\documentclass[a4paper,10pt]{article}
\usepackage[text={7in,9in},centering]{geometry}
\usepackage[utf8]{inputenc}
\usepackage[usenames]{color}

\newcommand{\cred}[1]{{\color{red} #1}}

\definecolor{brown}{rgb}{0.3,0.2,0}

\usepackage{graphicx,bm,amssymb,amsmath}
\usepackage{ulem} 

\newcommand{\brho} {\mbox {\boldmath $\rho$}}

\newcommand{\bb}{{\bf b}}

\newcommand{\bh}{{\bf h}}

\newcommand{\bj}{{\bf j}}

\newcommand{\bl}{{\bf l}}

\newcommand{\bq}{{\bf q}}
\newcommand{\br}{{\bf r}}

\newcommand{\bx}{{\bf x}}
\newcommand{\by}{{\bf y}}
\newcommand{\bz}{{\bf z}}

\newcommand{\bH}{{\bf H}}

\newcommand{\bJ}{{\bf J}}

\newcommand{\bS}{{\bf S}}
\newcommand{\bT}{{\bf T}}

\newcommand{\cD}{{\cal D}}

\newcommand{\cF}{{\cal F}}

\newcommand{\cH}{{\cal H}}

\newcommand{\cN}{{\cal N}}

\newcommand{\cZ}{{\cal Z}}

\newcommand{\I}{\mathbb{I}}

\newcommand{\N}{\mathbb{N}}
\newcommand{\R}{\mathbb{R}}

\newcommand{\Z}{\mathbb{Z}}

\def\tensor#1{\protect\@ontopof{#1}{\leftrightarrow}{1.15}\mathord{\box2}}

\DeclareMathOperator{\Tr}{Tr}

\title{The self-adjoint toroidal dipole operator in nanostructures}
\author{
	Mircea Dolineanu\thanks{Institutul National de Cercetare-Dezvoltare pentru Fizica si Inginerie Nucleara Horia Hulubei, University of Bucharest, Faculty of Physics, mircea.dolineanu@theory.nipne.ro},\
	Amanda Teodora Preda\thanks{Institutul National de Cercetare-Dezvoltare pentru Fizica si Inginerie Nucleara Horia Hulubei, University of Bucharest, Faculty of Physics, amanda.preda96@gmail.com} \ and
	Drago\c s-Victor Anghel\thanks{Institutul National de Cercetare-Dezvoltare pentru Fizica si Inginerie Nucleara Horia Hulubei, dragos@theory.nipne.ro; \textit{corresponding author}}
}

\begin{document}

\maketitle

\begin{abstract}
The parity violation in nuclear reactions led to the discovery of the new class of toroidal multipoles.
Since then, it was observed that toroidal multipoles are present in the electromagnetic structure of systems at all scales, from elementary particles, to solid state systems and metamaterials.
The toroidal dipole $\bT$ (the lowest order multipole) is the most common.
In quantum systems, this corresponds to the toroidal dipole operator $\hat{\bT}$, with the projections $\hat{T}_i$ ($i=1,2,3$) on the coordinate axes.
Here we analyze a quantum particle in a system with cylindrical symmetry, which is a typical system in which toroidal moments appear.
We find the expressions for the Hamiltonian, momenta, and toroidal dipole operators in adequate curvilinear coordinates, which allow us to find analytical expressions for the eigenfunctions of the momentum operators.
While the toroidal dipole is hermitian, it is not self-adjoint, but in the new set of coordinates the operator $\hat{T}_3$ splits into two components, one of which is (only) hermitian, whereas the other one is self-adjoint.
The self-adjoint component is the one which is physically significant and represents an observable.
Furthermore, we numerically diagonalize the Hamiltonian and the toroidal dipole operator and find their eigenfunctions and eigenvalues.
We write the partition function and calculate the thermodynamic quantities for a system of ideal particles on a torus.
Beside proving that the toroidal dipole is self-adjoint and therefore an observable (a finding of fundamental relevance) such systems open up the possibility of making metamaterials which exploit the quantization and the quantum properties of the toroidal dipoles.

{\bf Keywords:} toroidal dipole operator; quantum observables; nano-systems; metamaterials.
\end{abstract}

\section{Introduction} \label{sec_intro}

\begin{figure}
	\centering
	\includegraphics[height=4 cm]{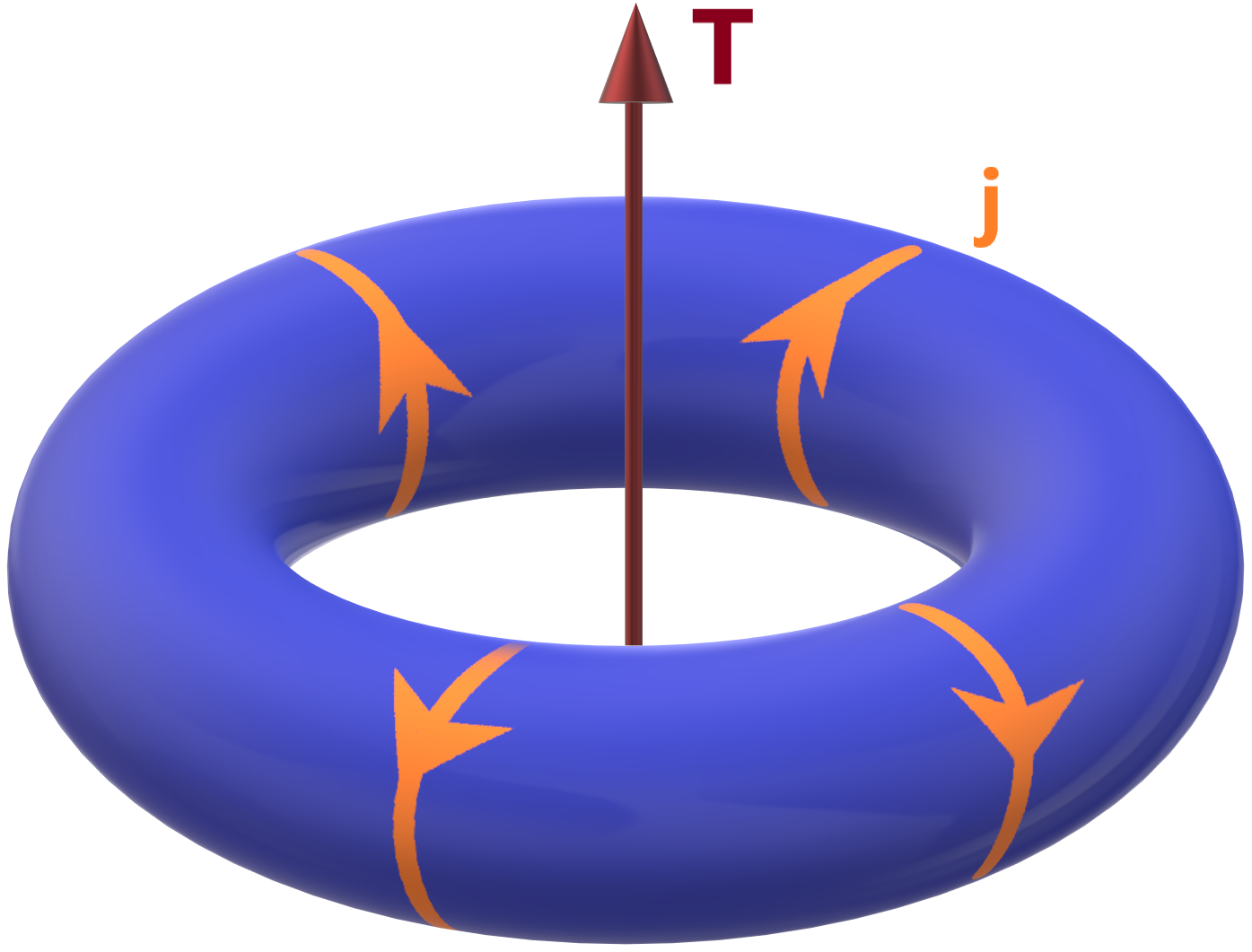}
%	\hspace{5mm}
%	\includegraphics[height=4 cm]{pics/Magnetic_moment.png}
	\caption{A toroidal dipole $\bT$ generated by the poloidal currents $\bj$ on a toroidal solenoid.
	}
	\label{fig_tor}
\end{figure}

In 1957, Zeldovich introduced a new type of electromagnetic interaction, in order to explain the parity nonconservation in $\beta$-decays~\cite{SovPhysJETP.6.1184.1958.Zeldovich}:
\begin{equation}
	\hat{H}_{\beta} \sim \bS \cdot \bJ^{ext} = \bS \cdot (\nabla \times H^{ext}) \label{H_Zeldovich}
\end{equation}
(where $\bS$ is the spin of the particle, $\bJ^{ext}$ is the external current, and $\bH^{ext}$ is the external magnetic field). Since neither electric nor magnetic multipoles lead to an interaction of the type~(\ref{H_Zeldovich}), he introduced for the first time the notion of "anapole", a type of distribution that he intuitively  described as a toroidal solenoid--a wire solenoid curved into a torus, as in Fig.~\ref{fig_tor}~(a).
A magnetic field is produced inside this distribution, due to the static toroidal current, while the electric field everywhere is zero.
In the context of classical electrodynamics, Dubovik and Cheshkov proposed in the late 1960s that a whole new multipole family has to be introduced to complement the electric and magnetic moments ~\cite{Sov.24.1965.Dubovik, SovJ.5.318.1974.Dubovik}.
They reached the conclusion that a dynamic nonradiating anapole can only be achieved if toroidal moments are taken into consideration. Fundamental symmetry considerations also explain the need for these new moments.
For example, the electric dipole moment is odd under spatial inversion and even under time reversal, while the magnetic dipole moment is odd under time reversal and even under spatial inversion. For a complete picture, one needs to include dipole moments which are even or odd under both transformations and these are the axial and polar toroidal dipole moments ~\cite{nanz:book.2016toroidal}.
It has also been realized that the toroidal moments and anapole configurations represent an important tool to describe the properties of systems at all scales, from particle physics to the physics of macroscopic systems and this new field expanded rapidly ~\cite{NatureMat.15.263.2016.Papasimakis, AIPConfProc.477.14.1999.Flambaum}.
For example, it was observed that particles that may be good candidates for dark matter constituents can only have toroidal moments ~\cite{PhysLettB.722.341.2013.Ho, NuclPhysB.907.1.2016.Cabral} and the CPT invariance alone requires that the electromagnetic structure of Majorana fermions consists only of toroidal moments~\cite{PhysRevD.32.1266.Radescu,ModPhysA.13.5257.1998.Dubovik}.

In solid state physics, toroidal ordering was first studied theoretically by Charles Kittel ~\cite{PhysRev.70.965.1946.Kittel}  and it indicates the existence of a new type of magnetoelectric effect. In the framework of condensed matter, toroidal moments are linked to another kind of order parameter known as toroidization  (toroidal polarization).
Media that exhibit macroscopic toroidization are called ferrotoroids and since they are expected to have promising technological applications (for example in data storage), it is crucial to investigate whether ferrotoroidicity is on an equal footing with the other feroic states (ferroelectric, ferromagnetic)~\cite{Physics.2.20.2009.Khomskii, PU.55.557.2012.Pyatakov, JExpTPL.52.161.1990.Tolstoi, NewJPhys.9.95.2007.Fedotov, PhysRevB.84.094421.2011.Toledano, PhysRevB.101.2020.Shimada, JexpThPhys.87.146.1998.Popov, NatNano.14.141.2019.Lehmann}.

Recently, anapole states have generated tremendous interest in the field of nanophotonics, optics, and metamaterials, and have great potential for a wide range of applications like lasers, sensing, and nonscattering objects that may be used for their cloaking behavior~\cite{Nanotechnology.30.2019.Yang, Comm.Pres.2.10.2019.Savinov, Nanophotonics.7.2017.Talebi, LasPhotRev.13.1800266.2019.Gurvitz}.
It has even been proposed that since anapole states interact weakly with electomagnetic fields, they could be used to protect qubits from environmental disturbance~\cite{ScRep.5.2016.Zagoskin}.

To analyze, manipulate, or engineer systems with toroidal properties at small scales (e.g., particles, molecules, nano-systems, metamaterials), one needs a quantum description of the toroidal moments.
The toroidal dipole operator was introduced in Ref.~\cite{AnnPhys.209.13.1991.Costescu}, whereas its eigenvalues and  eigenfunctions were found in~\cite{JPA30.3515.1997.Anghel}.
It was also shown that the toroidal dipole operator, when defined on the whole space, is hypermaximal (that is, it has several self-adjoint extensions)~\cite{JPA30.3515.1997.Anghel}, whereas on a finite definition domain, the Dirichlet boundary conditions impose that the operator is hermitian, but not self-adjoint.
%Here we solve this problem by introducing a new operator which has the same eigenvalues as the toroidal dipole operator, but, because it is defined on closed paths, satisfies periodic boundary conditions, which makes it self-adjoint.

In this paper, we analyze the energy and the toroidal dipole of a quantum particle on a quasi-two-dimensional (quasi-2D) domain folded into a shape with cylindrical symmetry, similar to a torus, but of a general cross-section.
We write the momentum, the Hamiltonian, and the projection $\hat{T}_3$ of toroidal dipole operators in the curved coordinates corresponding to such a domain.
We split $\hat{T}_3$ into two components, one of which is hermitian (but not self-adjoint) and the other one is self-adjoint.
By this, we eventually show for the first time that the toroidal dipole represents a quantum observable.

We particularize the system to a torus (made of a quasi-2D sheet and of circular cross-section) and calculate the eigenvalues and eigenfunctions of the studied operators.
Further, we show that while the Hamiltonian eigenstates of lower energies have zero toroidal dipole, at higher energies there is a crossover above which the toroidal dipole of the particle deviates from zero with an absolute value which increases with energy.
In the energy range above the crossover, the momentum, energy, and toroidal dipole eigenstates are (almost) identical.

Finally, we apply the formalism to calculate the thermodynamics of ideal bosons and fermions in such quasi-2D domains.
While in the high temperature limit the thermodynamics is similar to the thermodynamics of a 2D system, in the low temperature limit, the system of ideal fermions exhibit a ``toroidization'' property, due to the crossover between the energy domain of eigenstates with zero toroidal dipole expectation value to the domain with finite expectation values.

The paper is organized as follows.
In Section~\ref{sec_tor_op} we introduce the toroidal dipole operator and its basic properties.
In Section~\ref{sec_CSCO} we introduce the curvilinear system of coordinates, we write the momentum, Hamiltonian, and toroidal dipole operators in terms of them, and we give the general recipe to calculate the eigenvalues and eigenfunction.
In Section~\ref{self_adj_T3} we prove the self-adjointness of one of the components of $\hat{T}_3$, showing in this way that the toroidal dipole is an observable.
In Section~\ref{sec_eign_val_tor} we particularize the formalism to a torus and explicitly calculate the eigenvalues and eigenfunctions for all the relevant operators.
In Section~\ref{sec_thermo} we calculate the thermodynamics of a system of identical particles on a torus and in Section~\ref{sec_conclusions} we draw the conclusions.
In Appendix~\ref{sec_app_herm} we explicitly prove the hermiticity of the toroidal dipole operator in the curvilinear coordinates and introduce some notations necessary in the main body of the article.

%\section{Operators that commute with $\hat{T}_3$ and $\hat{L}_3$} \label{sec_CSCO}
\section{The toroidal diople of the particle} \label{sec_tor_op}

\begin{figure}
	\centering
	\includegraphics[height=5cm, keepaspectratio=true]{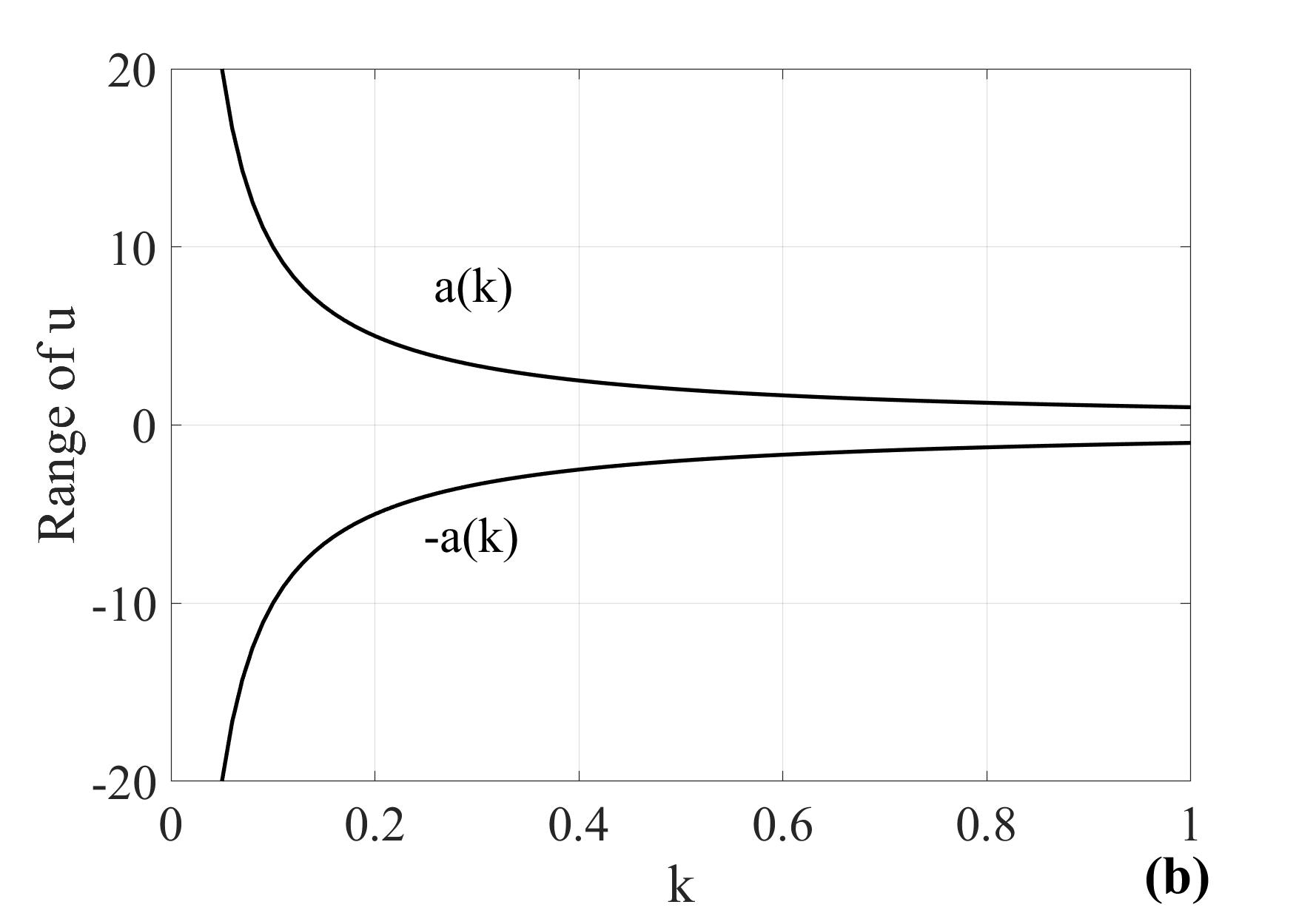}
	\caption{The ``natural coordinate'' $u$ takes values in the interval $u\in(-a(k), a(k))$, for any $k$.
	}
	\label{fig_range_u}
\end{figure}

The toroidal dipole is the lowest order toroidal moment.
It is a polar vector and, for a current distribution $\bj(\br)$, has the expression~\cite{PhysRep.187.145.1990.Dubovik}
\begin{equation} \label{def_Ti}
	\bT = \frac{1}{10} \int_V \Big[\br (\br \cdot \bj) - 2 r^2 \bj \Big] d^3\br ,
\end{equation}
where $r \equiv |\br|$.
%The component along the $z$ axis is
%%
%\begin{equation} \label{def_T3}
	%	T_3 = \frac{1}{10} \int_V \Big[r_3 r_k - 2 r^2 \delta_{3k} \Big] j_k d^3\br
	%	\equiv \int_V \bT^{(3)} \bj d^3\br ,
	%	\quad {\rm where} \quad
	%	T^{(3)}_k \equiv \frac{r_3 r_k - 2 r^2 \delta_{3k}}{10}
	%\end{equation}
%%
For quantum systems, the toroidal dipole~(\ref{def_Ti}) corresponds to the operator $\hat{\bT}$, of components~\cite{AnnPhys.209.13.1991.Costescu, JPA30.3515.1997.Anghel}
\begin{equation} \label{def_Ti_op}
	\hat T_i \equiv \frac{1}{10 m_p} \sum_{j=1}^{3} \left(x_i x_j - 2 r^2 \delta_{ij} \right) \hat p_j
\end{equation}
in Cartesian coordinates, where we used the notations $(x_1,x_2,x_3) \equiv (x,y,z)$ for the components of the position vector $\br$, $\hat p_j \equiv - i\hbar \partial/\partial x_j$ is the momentum operator along the $j$ axis, and $m_p$ is the mass of the particle.
%If $\hat{L}_j$ is the projection of the angular momentum operator on the $j$ axis, we have the commutation relations~\cite{JPA30.3515.1997.Anghel}
%%
%\begin{equation} \label{commut_TiLj}
	%[\hat T_i, \hat L_j] = i \hbar \epsilon_{ijk} \hat T_k .
	%\end{equation}
%%
%In Ref.~\cite{JPA30.3515.1997.Anghel} it was shown that in the $\R^3$ space $\hat T_i$ are (non-commuting) hypermaximal operators.
The projection $\hat{T}_3$ has a simpler form in cylindrical coordinates~\cite{JPA30.3515.1997.Anghel},
\begin{equation} \label{def_T3_op_cyl}
	\hat T_3 = \frac{-i\hbar}{10 m_p} \left[ z\rho \frac{\partial}{\partial \rho} - \left( 2\rho^2 + z^2 \right) \frac{\partial}{\partial z} \right] ,
	%\equiv  -i\hbar \frac{\partial}{\partial u} ,
\end{equation}
where $\rho \equiv \sqrt{x^2 + y^2}$ and $r = \sqrt{\rho^2 + z^2}$.
One may define the vector field $\bT_3 \equiv z\rho \hat{\bf\rho} - (2\rho^2+z^2) \hat{\bz}$ and the ``natural coordinates'' $(k,u,\phi)$ of $\hat T_3$, such that
\begin{equation} \label{def_T3_cn}
	\hat T_3 \equiv  -i \hbar \frac{\partial}{\partial u} ,
\end{equation}
(notice that $u$ was introduced in~\cite{JPA30.3515.1997.Anghel} by a different definition, namely $\hat{T}_3 = - [i\hbar / (10m_pc)] \partial / \partial u$).
We observe that the vector field $\bT_3$ is tangent in any point to the curves that define the coordinate $u$.
Furthermore, the definition~(\ref{def_T3_cn}) leads to~\cite{JPA30.3515.1997.Anghel}
\begin{equation}
	k \equiv \left[ \rho^2 \left( z^2 + \rho^2 \right) \right]^{1/4} \equiv \sqrt{\rho r} \quad {\rm and} \quad
	u \equiv f_u(k,z) = - 10 m_p \int_0^z \frac{dt}{\sqrt{t^4 + 4 k^4}} = \pm 10 m_p \int_\rho^k \frac{dt}{\sqrt{-t^4 + k^4}} . \label{def_nat_set}
\end{equation}
%
% (where we introduced the notation $f_u(k,z)$).
The variable $u$ takes values in a finite interval $(-a(k), a(k))$, where~\cite{JPA30.3515.1997.Anghel}
\begin{equation} \label{def_ak}
	a(k) = \frac{10 m_p}{4k} \frac{\Gamma\left(\frac{1}{4}\right) \Gamma\left(\frac{1}{2}\right)}{\Gamma\left(\frac{3}{4}\right)}
	\equiv \frac{10 m_p}{k} C_a
	%    \approx \frac{1.31103}{k} ,
\end{equation}
and $C_a \approx 1.31103$ (see Fig.~\ref{fig_range_u}). \cred{\AA}
The end points $u\to-a(k)$ and $u\to a(k)$ (at fixed $k$) correspond to the points $(z\to \infty, \rho \to 0)$ and $(z\to -\infty, \rho \to 0)$, respectively.

\section{The quantum particle in a system with thin walls and cylindrical symmetry} \label{sec_CSCO}

\begin{figure}[t]
	\centering
	\includegraphics[width=18cm]{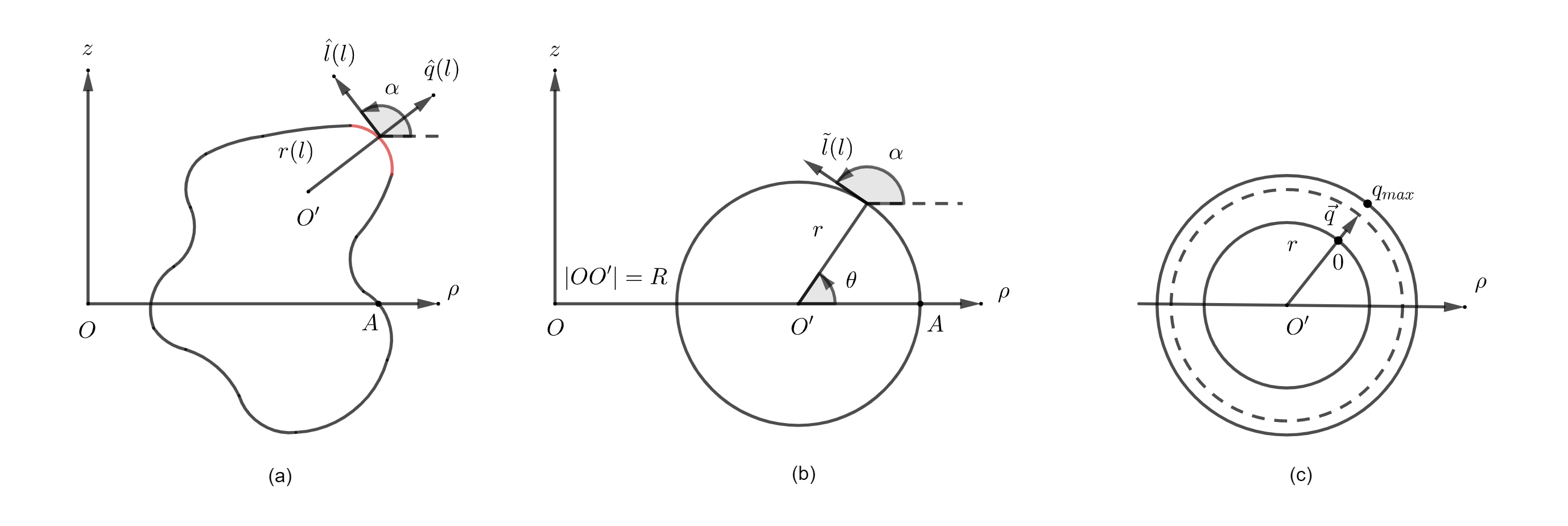}
	\caption{
	The cross-sections of two geometries with cylindrical symmetry: (a) a general shape and (b) a torus.
	The cross-section of the torus is centered in $O'$, $|OO'| \equiv R$ is the major radius, and $r$ is the minor radius.
	In both, (a) and (b), $A$ is the origin of the coordinate $l$.
	The local basis is $(\hat{\bl}(l),\hat{\bq}(l))$ and the angle $\alpha$ is defined such that $\hat{\bl} \cdot \hat{\brho} = \cos\alpha, \hat{\bl} \cdot \hat{\bz} = \sin\alpha, \hat{\bq} \cdot \hat{\brho} = \sin\alpha, \hat{\bq} \cdot \hat{\bz} = -\cos\alpha$.
	In (a), the local curvature is centered at $O'(l)$ and has a radius $r(l)$
	In (b), $O'$ does not depend on $l$ and $\theta = \alpha - \pi/2$.
	The finite thickness of the sheet is shown in (c) and is measured by the coordinate $q$ ($0\le q \le q_{max}$)
	}
	\label{fig_contour}
\end{figure}

We analyze a quantum particle on a quasi-two-dimensional (2D) sheet of material $\Lambda$ with cylindrical symmetry, similar to the one presented in Fig.~\ref{fig_tor}, but of arbitrary cross-section, as exemplified in Fig.~\ref{fig_contour}~(a).
Following the procedure outlined in~\cite{PhysLettA.380.1985.2016.Shikakhwa, PhysLettA.380.2876.2016.Shikakhwa}, we define the set of curvilinear orthogonal coordinates $(l, q, \phi)$, where $\phi$ is the azimuthal angle, $q$ is the coordinate perpendicular to the surface in any point (Fig.~\ref{fig_contour}~c), and $l$ is the coordinate along the cross-section, starting and ending in $A$, as shown in Fig.~\ref{fig_contour}~(a) and (b).
Although the system is considered quasi-2D, the coordinate $q$ takes finite values and varies from $0$ (the interior surface of the system) to $q_{max}$ (the exterior surface), where $q_{max}$ is much smaller than all the other dimensions of the system (see Fig.~\ref{fig_contour}~c).
In such a case, the ``point'' $A$ from the 2D representation of $\Lambda$ corresponds actually to $l=0$ and $l=L$ (the maximum value of $l$) for any $q\in[0,q_{max}]$.

\begin{figure}[t]
	\centering
	\includegraphics[height=5cm]{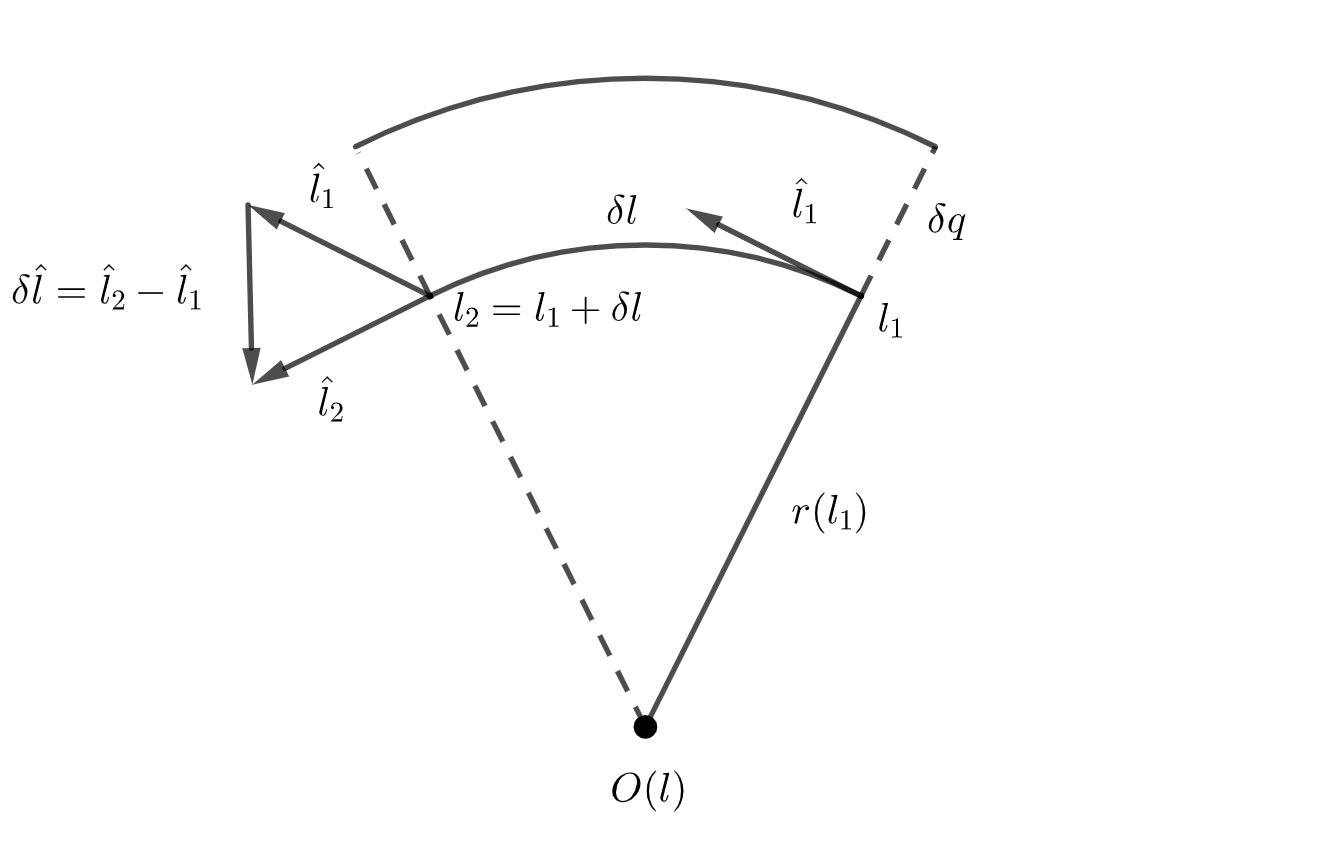}
	\caption{
	The method for calculating $\partial \hat{\bl}/ \partial l$ and $\partial h_l/\partial q$: the variation $\delta l$ of $l$ leads to the variation $\delta \hat{\bl}$ of the direction $\hat{\bl}$, whereas the variation $\delta q$ of $q$ causes a variation $\delta h_l$ of $h_l$.
	Then, $|\partial \hat{\bl}/ \partial l| = 1/|r(l)|$, $\partial h_l/\partial q = 1/r(l)$, where $|r(l)|$ is the radius of the local curvature, centered at $O(l)$.
	The sign of $r(l)$ is chosen such that $r(l) \equiv - (\delta \hat{\bl} \cdot \hat{\bq}) |r(l)|$, assuming that $\hat{\bq}$ points towards the outside of the system for any $l$ (see also Fig.~\ref{fig_contour}).
	}
	\label{fig_angles}
\end{figure}

If $(\hat{\bx}, \hat{\by}, \hat{\bz}) \equiv (\hat{\bx}_1, \hat{\bx}_2, \hat{\bx}_3)$ is the Cartesian coordinates system and the position vector is $\br \equiv \sum_i x_i \hat{\bx}_i$, then we define the set of reciprocally orthogonal vectors
\begin{equation}
	\bh_l \equiv \frac{\partial \br}{\partial l}, \quad
	\bh_q \equiv \frac{\partial \br}{\partial q}, \quad
	\bh_\phi \equiv \frac{\partial \br}{\partial \phi}
	= -y \hat{\bx} + x \hat{\by} ,
	\label{def_hs}
\end{equation}
and the local basis
\begin{equation}
	\bb_l \equiv \frac{\bh_l}{h_l}, \quad
	\bb_q \equiv \frac{\bh_q}{h_q}, \quad
	\bb_\phi \equiv \frac{\bh_\phi}{h_\phi} ,
	\quad {\rm where} \quad
	h_l \equiv |\bh_l|, \quad
	h_q \equiv |\bh_q|, \quad
	h_\phi \equiv |\bh_\phi| = \sqrt{x^2 + y^2} .
	\label{def_bhs}
\end{equation}
The partial derivatives may be written in the local basis as
\begin{equation}
	\left(\begin{array}{c}
		\frac{\partial}{\partial x} \\
		\frac{\partial}{\partial y} \\
		\frac{\partial}{\partial z}
	\end{array}\right)
	=
	[J_{l,q,\phi}]
	\left(\begin{array}{c}
		\frac{\partial}{\partial l} \\
		\frac{\partial}{\partial q} \\
		\frac{\partial}{\partial \phi}
	\end{array}\right) ,
	\quad
	\left(\begin{array}{c}
		\frac{\partial}{\partial l} \\
		\frac{\partial}{\partial q} \\
		\frac{\partial}{\partial \phi}
	\end{array}\right)
	=
	[J_{l,q,\phi}]^{-1}
	\left(\begin{array}{c}
		\frac{\partial}{\partial x} \\
		\frac{\partial}{\partial y} \\
		\frac{\partial}{\partial z}
	\end{array}\right) ,
	\quad {\rm where} \quad
	[J_{l,q,\phi}] \equiv
	\left(\begin{array}{ccc}
		\frac{\partial l}{\partial x} & \frac{\partial q}{\partial x} & \frac{\partial \phi}{\partial x} \\
		\frac{\partial l}{\partial y} & \frac{\partial q}{\partial y} & \frac{\partial \phi}{\partial y} \\
		\frac{\partial l}{\partial z} & \frac{\partial q}{\partial z} & \frac{\partial \phi}{\partial z}
	\end{array}\right) .
	\label{ders_part_change}
\end{equation}
%\begin{equation}
%	\left(\begin{array}{c}
%		\frac{\partial}{\partial x} \\
%		\frac{\partial}{\partial z} \\
%		\frac{\partial}{\partial z}
%	\end{array}\right)
%	\equiv
%	\left(\begin{array}{ccc}
%		\frac{\partial l}{\partial x} & \frac{\partial q}{\partial x} & \frac{\partial \phi}{\partial x} \\
%		\frac{\partial l}{\partial y} & \frac{\partial q}{\partial y} & \frac{\partial \phi}{\partial y} \\
%		\frac{\partial l}{\partial z} & \frac{\partial q}{\partial z} & \frac{\partial \phi}{\partial z}
%	\end{array}\right)
%	\left(\begin{array}{c}
%	\frac{\partial}{\partial l} \\
%	\frac{\partial}{\partial q} \\
%	\frac{\partial}{\partial \phi}
%	\end{array}\right) ,
%%
%	\quad {\rm whereas} \quad
%%
%\left(\begin{array}{c}
%	\frac{\partial}{\partial l} \\
%	\frac{\partial}{\partial q} \\
%	\frac{\partial}{\partial \phi}
%\end{array}\right)
%\equiv
%\left(\begin{array}{ccc}
%	\frac{\partial x}{\partial l} & \frac{\partial y}{\partial l} & \frac{\partial z}{\partial l} \\
%	\frac{\partial x}{\partial q} & \frac{\partial y}{\partial q} & \frac{\partial z}{\partial q} \\
%	\frac{\partial x}{\partial \phi} & \frac{\partial y}{\partial \phi} & \frac{\partial z}{\partial \phi}
%\end{array}\right)
% \left(\begin{array}{c}
% 	\frac{\partial}{\partial x} \\
% 	\frac{\partial}{\partial z} \\
% 	\frac{\partial}{\partial z}
% \end{array}\right) .
%\label{ders_part_change}
%\end{equation}
%
The volume element in the curvilinear coordinates is
\begin{equation}
	d^3\br \equiv dx\,dy\,dz \equiv |\det [J_{l,q,\phi}]| dl\,dq\,d\phi = \rho h_l h_q dl\,dq\,d\phi .
	\label{dV_ccoord}
\end{equation}
We choose the coordinates such that
\begin{equation}
%	h_q = 1, \quad h_l \equiv 1 + q h'(l), \quad {\rm where} \quad h'(l) = \frac{\partial h_l}{\partial q} .
	h_q = 1, \quad {\rm so} \quad h_l = 1 + q\frac{\partial h_l}{\partial q} = 1 + \frac{q}{r(l)}, \quad |\det [J_{l,q,\phi}]| = \rho h_l ,
	\label{hl_lin}
\end{equation}
and $r(l)$ is the radius of the local curvature, centered in $O'(l)$ (see Figs.~\ref{fig_contour} and~\ref{fig_angles}).
We assume $r(l) \ne 0$, for any $l \in [0, L]$ (which implies also that $r(0) = r(L) \ne 0$).

Whereas $\hat{L}_z \equiv -i\hbar \partial/\partial \phi$ is the projection of the angular momentum on the $z$ axis and is self-adjoint due to the periodic boundary conditions in $\phi$, the operators $-i\hbar\partial/\partial l$ and $-i\hbar\partial/\partial q$ may not be self-adjoint or even hermitian.
The hermitian operators associated to $-i\hbar\partial/\partial l$ and $-i\hbar\partial/\partial q$ are the momentum operator projection on the local $l$ and $q$ axes~\cite{PhysLettA.380.1985.2016.Shikakhwa}
\begin{equation} \label{adjoint_dl_dq}
	\hat{p}^{(\bl)} \equiv -i\hbar \left(\frac{1}{h_l}\frac{\partial}{\partial l} + \frac{1}{2\rho h_l}\frac{\partial \rho}{\partial l}\right)
	\quad {\rm and} \quad
	\hat{p}^{(\bq)} \equiv -i\hbar \left[\frac{\partial}{\partial q} + \frac{1}{2\rho h_l}\frac{\partial (\rho h_l)}{\partial q}\right] .
\end{equation}
For the choices~(\ref{hl_lin}) we obtain
\begin{subequations} \label{adjoint_dl_dq_special}
\begin{eqnarray}
%	\hat{p}^{(\bl)} &=& -i\hbar \left\{\frac{\partial}{\partial l} + \frac{h_l \hat{\brho} \cdot \hat{\bl}}{2\rho} + \frac{q}{2 h_l}\frac{d}{d l}\left(\frac{1}{r(l)}\right) \right\}
	\hat{p}^{(\bl)} &=& -i\hbar \left(\frac{1}{h_l}\frac{\partial}{\partial l} + \frac{\hat{\brho} \cdot \hat{\bl}}{2\rho} \right)
	\label{adjoint_dl} \\
	\hat{p}^{(\bq)} &=& -i\hbar \left(\frac{\partial}{\partial q} + \frac{\hat{\brho} \cdot \hat{\bq}}{2\rho} + \frac{1}{2 h_l r} \right) .
	\label{adjoint_dq}
%	\approx -i\hbar \left\{\frac{\partial}{\partial q} + \frac{\hat{\brho} \cdot \hat{\bq}}{2\rho} + \frac{h'(l)[1-h'(l) q]}{2} \right\} .
\end{eqnarray}
\end{subequations}
Since the points $l=0$ and $l=L$ coincide for all $q$ and $\phi$, the wavefunction satisfies periodic boundary conditions in $l$.

%Imposing periodic boundary conditions in the variable $l$, since the points $l=0$ and $l=L$ coincide for all $q,\phi$, the operator $\hat{p}^{(\bl)}$ is self-adjoint, whereas $\hat{p}^{(\bq)}$ is not, since at boundary in the variable $q$ the wavefunctions are zero.

Similarly, the toroidal dipole operator (\ref{def_T3_op_cyl}) may be written
%%
%\begin{equation} \label{def_Ti_op_crv}
%\hat T_i \equiv \frac{1}{10 m_p} \sum_{j=1}^{3} \left(x_i x_j - 2 r^2 \delta_{ij} \right) \hat p_j
%\end{equation}
%%
%
\begin{eqnarray} % \label{def_T3_op_cyl_crv}
\hat T_3
%&=& \frac{-i\hbar}{10 m_p} \left[ z\rho \left( \frac{\partial l}{\partial \rho} \frac{\partial}{\partial l} + \frac{\partial q}{\partial \rho} \frac{\partial}{\partial q} \right) - \left( 2\rho^2 + z^2 \right) \left( \frac{\partial l}{\partial z} \frac{\partial}{\partial l} + \frac{\partial q}{\partial z} \frac{\partial}{\partial q} \right) \right]
%\nonumber \\
%
%&=& \frac{-i\hbar}{10 m_p} \left\{
%\left[
%z\rho \left(\frac{\partial l}{\partial \rho}\right)_z - \left( 2\rho^2 + z^2 \right) \left(\frac{\partial l}{\partial z}\right)_\rho \right] \frac{\partial}{\partial l}
%+
%\left[
%z \rho \left(\frac{\partial q}{\partial \rho}\right)_z - \left( 2\rho^2 + z^2 \right) \left(\frac{\partial q}{\partial z}\right)_\rho \right] \frac{\partial}{\partial q}
%\right\} \nonumber \\
%
&=& \frac{-i\hbar}{10 m_p} \left[
	z\rho \,\hat{\brho} - \left( 2\rho^2 + z^2 \right) \,\hat{\bz} \right] \cdot \left(
\frac{\hat{\bl}}{h_l} \frac{\partial}{\partial l}
+ \hat{\bq}\frac{\partial}{\partial q}
\right) .
\label{T_3tot}
\end{eqnarray}
One can easily check from Eqs.~(\ref{def_Ti_op}) and (\ref{def_T3_op_cyl}) that the operators $\hat{T}_i$, with periodic or Dirichlet boundary conditions, are hermitian~\cite{JPA30.3515.1997.Anghel}.
This implies that $\hat{T}_3$ is hermitian also in the coordinates $(l,q,\phi)$ (Eq.~\ref{T_3tot}), but in order to exemplify the action of the partial derivatives on different coordinates and to introduce some notations, we explicitly prove the hermiticity in the coordinates $(l,q,\phi)$ in Appendix~\ref{sec_app_herm}.

The Hamiltonian was calculated in Refs.~\cite{PhysLettA.380.1985.2016.Shikakhwa, PhysLettA.380.2876.2016.Shikakhwa}, so (without taking into account the fields) we use the expression
\begin{subequations} \label{defs_H_SOC}
	\begin{eqnarray}
	\hat{\cH}
	&=& \frac{-\hbar^2}{2m_p} \Delta
	= \frac{-\hbar^2}{2m_p} \frac{1}{h_lh_qh_\phi}
	\left( \frac{\partial}{\partial l} \frac{h_qh_\phi}{h_l} \frac{\partial}{\partial l}
	+ \frac{\partial}{\partial q} \frac{h_lh_\phi}{h_q} \frac{\partial}{\partial q }
	+ \frac{\partial}{\partial \phi} \frac{h_lh_q}{h_\phi} \frac{\partial}{\partial \phi} \right)
	\nonumber \\
	%
%	&=&
%	\frac{-\hbar^2}{2m_p} \frac{1}{\rho h_l}
%	\left[ \frac{\partial}{\partial l} \frac{\rho}{h_l} \frac{\partial}{\partial l}
%	+ \frac{\partial}{\partial q} \rho h_l \frac{\partial}{\partial q }
%	+ \frac{h_l}{\rho} \frac{\partial^2}{\partial \phi^2} \right]
%	\nonumber \\
%	%
%	&=&
%	\frac{-\hbar^2}{2m_p} \frac{1}{\rho h_l}
%	\left[ \left(\hat{\brho} \cdot \hat{\bl} - \frac{\rho q}{h_l^2} \frac{dh'}{dl} \right) \frac{\partial}{\partial l}
%	+ \frac{\rho}{h_l} \frac{\partial^2}{\partial l^2}
%	+ \left( h_l \,\hat{\brho} \cdot \hat{\bq} + \frac{\rho}{r(l)} \right) \frac{\partial}{\partial q }
%	+ \rho h_l \frac{\partial^2}{\partial q^2 }
%	+ \frac{h_l}{\rho} \frac{\partial^2}{\partial \phi^2} \right]
%	\nonumber \\
	%
	&=&
	\frac{-\hbar^2}{2m_p} \left\{ \frac{1}{\rho h_l}
	\left[ \left(\hat{\brho} \cdot \hat{\bl} - \frac{\rho q}{h_l^2} \frac{d}{dl}\left(\frac{1}{r}\right) \right) \frac{\partial}{\partial l}
	+ \frac{\rho }{h_l}\frac{\partial^2}{\partial l^2}
	+ \frac{h_l}{\rho} \frac{\partial^2}{\partial \phi^2} \right]
	+ M^2 - K \right\}
	+ \frac{(\hat{p}^{(\bq)})^2}{2m_p}
	\label{def_H1}
	\end{eqnarray}
	where
	\begin{eqnarray}
	M^2-K
	&\equiv& - \frac{1}{2 h_lh_\phi}\frac{\partial^2(h_lh_\phi)}{\partial q^2} + \frac{1}{(2 h_lh_\phi)^2} \left(\frac{\partial(h_l h_\phi)}{\partial q}\right)^2
	= \frac{1}{4}\left(\frac{\hat{\brho} \cdot \hat{\bq}}{\rho} - \frac{1}{h_lr}\right)^2 .
	\label{def_Hc}
	\end{eqnarray}
\end{subequations}

\subsection{The operators for the quasi-two-dimensional quantum particle} \label{subsec_quasi2D}

Assuming that the system is tightly confined in the $q$ direction and $q_{max} \ll r(l)$, one may use the approximations $\rho(l,q) \approx \rho(l,0), z(l,q) \approx z(l,0)$, $r(l,q) \approx r(l,0)$ and simplify the expression of the Hamiltonian to
\begin{eqnarray}
\hat{\cH}_0 &\approx&
\frac{-\hbar^2}{2m_p} \left[ \frac{1}{\rho}
\left( \rho \frac{\partial^2}{\partial l^2}
+ \hat{\brho} \cdot \hat{\bl} \frac{\partial}{\partial l}
+ \frac{1}{\rho} \frac{\partial^2}{\partial \phi^2} \right)
+ \frac{1}{4}\left(\frac{\hat{\brho} \cdot \hat{\bq}}{\rho} - \frac{1}{r}\right)^2 \right]
+ \frac{(\hat{p}^{(\bq)})^2}{2m_p} ,
\label{def_H2D}
\end{eqnarray}
which may be split into three components
\begin{subequations} \label{def_H0_fact}
\begin{equation}
	\hat{\cH}_0 \equiv \hat{\cH}^{(l)} + \hat{\cH}^{(\phi)} + \hat{\cH}_{(q)} , \quad
	\hat{\cH}^{(l)} \equiv \frac{-\hbar^2}{2m_p} \left[ \frac{1}{\rho}
	\left( \rho \frac{\partial^2}{\partial l^2}
	+ \hat{\brho} \cdot \hat{\bl} \frac{\partial}{\partial l} \right)
	+ \frac{1}{4}\left(\frac{\hat{\brho} \cdot \hat{\bq}}{\rho} - \frac{1}{r}\right)^2 \right] , \quad
	\hat{\cH}^{(\phi)} \equiv \frac{-\hbar^2}{2m_p} \frac{1}{\rho^2} \frac{\partial^2}{\partial \phi^2} , \quad
	\hat{\cH}^{(q)} \equiv \frac{(\hat{p}^{(\bq)})^2}{2m_p} ,
	 \label{def_H0}
\end{equation}
acting separately on the variables $l, \phi, q$.
So, its eigenfunctions may be written
\begin{equation}
	\Psi_{k_l, k_q, m} (l,q,\phi) \equiv \psi_{k_l} (l) \psi_{k_q} (q) \psi_{m} (\phi) , \quad
	\psi_{k_q} (q) \equiv \frac{\sin(k_q q)}{\sqrt{q_{max}}} , \quad
	\psi_{m}(\phi) \equiv \frac{e^{i m \phi}}{\sqrt{2\pi}} , \quad
	k_q = \frac{n_q \pi}{q_{max}}, \ n_q \in \N^+ , \ m \in \Z ,
\end{equation}
\end{subequations}
where $\N^+$ is the set of strictly positive integers and $\Z$ is the set of integers.
Since $\hat{\cH}_{(q)} \psi_{k_q}(q) = [n_q^2(\hbar\pi/q_{max})^2/(2m_p)] \psi_{k_q}(q)$, due to the tight confinement in the $q$ direction, the energies of the excited states (i.e., $n_q>1$) may be too high to be physically relevant and the system stays in the ground state, with $n_q = 1$.
Then, our problem becomes (quasi-)2D, in the variables $l$ and $\phi$, with the Hamiltonian
\begin{equation}
	\hat{\cH}^{(2D)} \equiv \hat{\cH}^{(l)} + \hat{\cH}^{(\phi)} .
	\label{H_2D}
\end{equation}

The confinement affects the action of all operators, including $\hat{T}_3$.
We use Eqs.~(\ref{calc_Tl_Tq}) from the Appendix and denote
\begin{subequations} \label{defs_t1_t2}
\begin{equation}
%\langle t^{(q)}_1 \rangle \equiv \frac{-i\hbar}{10 m_p} \int_{0}^{L} dl \int_0^{q_{max}} dq\, \rho h_l \left[ z\rho \, \hat{\brho} - \left( 2\rho^2 + z^2 \right) \,\hat{\bz} \right] \cdot \hat{\bq} \, \left(\Psi^* \frac{\partial}{\partial q} \Psi\right)
\left\langle t^{(q)}_1 \right\rangle \equiv \left\langle \Psi \left| \frac{-i\hbar}{10 m_p} \left[ z\rho \, \hat{\brho} - \left( 2\rho^2 + z^2 \right) \,\hat{\bz} \right] \cdot \hat{\bq} \,  \frac{\partial}{\partial q} \right| \Psi \right\rangle
\quad {\rm and} \quad
%\label{def_t1} \\
%
%&&
%\langle t_l \rangle \equiv \langle \Psi | \frac{T_l(l,q)}{h_l\rho} | \Psi \rangle ,
%\quad
\left\langle t^{(q)}_2 \right\rangle \equiv \left\langle \Psi \left| \frac{-i\hbar}{10 m_p} \frac{T_q(l,q)}{h_l\rho} \right| \Psi \right\rangle . \label{def_tlq}
\end{equation}
Derivating by parts, with Dirichlet boundary conditions in $q$, we observe that
\begin{equation}
	\langle t^{(q)}_1 \rangle = \langle t^{(q)}_1 \rangle^* + \langle t^{(q)}_2 \rangle ,
	\quad {\rm so} \quad
	\langle t^{(q)}_2 \rangle = \langle t^{(q)}_1 \rangle - \langle t^{(q)}_1 \rangle^* \in \I ,
	\quad \langle t^{(q)}_1 \rangle - \frac{\langle t^{(q)}_2 \rangle}{2} = \frac{\langle t^{(q)}_1 \rangle + \langle t^{(q)}_1 \rangle^*}{2} \in \R .
	\label{rels_t1_t2}
\end{equation}
\end{subequations}
Since the expectation value $\langle \Psi | \hat T_3 | \Psi \rangle$ is real ($\hat T_3$ is hermitian), using Eqs.~(\ref{defs_t1_t2}) we define two hermitian components of $\hat{T}_3$,
\begin{subequations} \label{defs_T3l_T3q}
\begin{eqnarray}
&& \hat{T}^{(l)}_3 \equiv - \frac{i\hbar}{10 m_p} \left\{\left[ z\rho \hat{\brho} - \left( 2\rho^2 + z^2 \right) \hat{\bz} \right] \cdot \frac{\hat{\bl}}{h_l} \frac{\partial}{\partial l}
+ \frac{T_l(l,q)}{2 \rho h_l} \right\} ,
\label{def_T3l} \\
&& \hat{T}^{(q)}_3 \equiv - \frac{i\hbar}{10 m_p} \left\{\left[ z\rho \hat{\brho} - \left( 2\rho^2 + z^2 \right) \hat{\bz} \right] \cdot \hat{\bq} \frac{\partial}{\partial q}
+ \frac{T_q(l,q)}{2 \rho h_l} \right\} ,
%\left\{ \rho \left[ z\rho \hat{\brho} - \left( 2\rho^2 + z^2 \right) \hat{\bz} \right] \cdot \hat{\bq} \right\} ,
\label{def_T3q}
\end{eqnarray}
such that
\begin{eqnarray}
&& \hat{T}_3 = \hat{T}^{(l)}_3 + \hat{T}^{(q)}_3 \label{T3_T3l_T3q}
\end{eqnarray}
\end{subequations}
(Eq.~\ref{Tl_Tq_0}).

\subsection{Eigenvalues and eigenvectors for the quasi-two-dimensional operators} \label{subsec_eigens}

The operator $\hat{p}^{(\bl)}$ %is self-adjoint and therefore it must
has a complete orthonormal set of eigenfunctions.
In the quasi-2D case we ignore the $q$ coordinate and from the expression~(\ref{adjoint_dl}) we obtain the eigenfunctions
\begin{subequations} \label{eigenfcts_pl_L3}
\begin{equation}
	\cF_{k_l}(l) \equiv \frac{1}{\sqrt{L}} \frac{e^{i l k_l}}{\sqrt{\rho(l)}} , \quad {\rm so} \quad
	\hat{p}^{(\bl)} \cF_{k_l}(l) = \hbar k_l \cF_{k_l}(l),
	\quad {\rm where} \quad
	k_l = \frac{2\pi n}{L}
	\quad {\rm and} \quad
	n = {\rm integer} .
	\label{ansatz}
\end{equation}
We can now introduce a basis in the $L^2(\Lambda)$ space of functions integrable in modulus square in the volume $\Lambda$, namely
\begin{equation} \label{q_cond}
	\cF^{(\Lambda)}_{k_l,m} \equiv \frac{1}{\sqrt{2\pi L}}\frac{e^{i (l k_l + m\phi)}}{\sqrt{\rho(l)}} ,
	\quad
	k_l = \frac{2\pi n}{L}
	\quad {\rm and} \quad
	n,m = {\rm integers} .
\end{equation}
One can easily check that
\begin{equation}
	\left\langle \cF^{(\Lambda)}_{k_l,m} | \cF^{(\Lambda)}_{k_l',m'} \right\rangle = \delta_{k_l,k_l'} \delta_{m,m'} \label{orthonormal}
\end{equation}
\end{subequations}
and the set $\{\cF^{(\Lambda)}_{k_l,m}(l,\phi)\}$ is complete in the Hilbert space of functions of coordinates $(l,\phi)$, integrable in modulus square.
For a thin torus $\rho(l) \gg L$ and one may use the approximation
\begin{equation}
	\cF^{(\Lambda)}_{k_l,m} \approx \frac{e^{i (l k_l + m\phi)}}{\sqrt{2\pi L \rho_0}} ,
\end{equation}
where $\rho_0$ is an approximation for $\rho(l)$ (e.g., the average) and $\rho_0, \rho(l) \gg L$ for any $0\le l < L$.

We calculate the matrix elements
\begin{subequations} \label{matr_elem_T3H_gen}
\begin{eqnarray}
	\langle \cF^{(\Lambda)}_{k_1,m_1} | \hat{T}_3^{(l)} | \cF^{(\Lambda)}_{k_2,m_2} \rangle &=& \delta_{m_1m_2} \langle \cF_{k_1} | \hat{T}_3^{(l)} | \cF_{k_2} \rangle
	\label{matr_elem_T3_gen} \\
	\langle \cF^{(\Lambda)}_{k_1,m_1} | \hat{\cH}^{(2D)} | \cF^{(\Lambda)}_{k_2,m_2} \rangle &=& \delta_{m_1m_2} \langle \cF_{k_1} | \hat{\cH}^{(2D)} | \cF_{k_2} \rangle ,
	\label{matr_elem_H_gen}
\end{eqnarray}
\end{subequations}
and, using them, we calculate the eigenvalues and eigenvectors of the operators $\hat{T}_3^{(l)}$ and $\hat{\cH}^{(2D)}$ from the  secular equation.
Notice that because of their hermiticity, the operators $\hat{\cH}^{(q)}$ and $\hat{T}_3^{(q)}$ may contribute only to the diagonal elements~(\ref{matr_elem_T3H_gen}), with quantities that do not depend on $k_l$ and $m$.
For this reason, their contributions will be neglected, since they only shift the eigenvalues.

\section{Self-adjointness of the operator $\hat{T}_3^{(l)}$} \label{self_adj_T3}

As we mentioned before, the operator $\hat{T}_3$ is hermitian, but not self-adjoint, due to the Dirichlet boundary conditions on the frontier of $\Lambda$.
Nevertheless, in Section~\ref{sec_CSCO} we split $\hat{T}_3$ into the hermitian components $\hat{T}_3^{(l)}$ and $\hat{T}_3^{(q)}$.
$\hat{T}_3^{(q)}$ is hermitian, but not self-adjoint, because the wavefunctions $\Psi(l,q,\phi)$ satisfy Dirichlet boundary conditions at $q = 0$ and  $q = q_{max}$.
On the other hand, $\Psi(l,q,\phi)$ is periodic in the variables $l$ and $\phi$ and, based on this property, we show that $\hat{T}_3^{(l)}$ is self-adjoint.

Let's denote by $L^{(2,l)}$ the Hilbert space of functions in the $l$ variable, which are integrable in modulus square.
We further denote by $\cD(\hat{T}_3^{(l)})$ and $\cD\left[\left({\hat{T}_3^{(l)}}\right)^\dagger\right]$ the definition domains of $\hat{T}_3^{(l)}$ and $\left({\hat{T}_3^{(l)}}\right)^\dagger$, respectively.
We define the derivative $dg/dl$ in a generalized sense by
\begin{equation}
	\int_{0}^{L} f(l) \frac{d g}{dl} \equiv f(L)g(L) - f(0)g(0) - \int_{0}^{L} g(l) \frac{d f}{dl}
	\label{def_der_gen}
\end{equation}
where $f, g$ are any functions in $L^{(2,l)}$, but $f$ is of class $C^1\left(L^{(2,l)}\right)$ (that is, it is derivable in any point and the derivative is continuous and integrable in modulus square).
If the derivative in the expression of $\hat{T}_3^{(l)}$ is interpreted in this generalized sense, then, $\cD\left({\hat{T}_3^{(l)}}\right)$ contains all the functions $f \in L^{(2,l)}$ that satisfy periodic boundary conditions and $\hat{T}_3^{(l)} f \in L^{(2,l)}$.
Further, $f\in \cD\left[\left({\hat{T}_3^{(l)}}\right)^\dagger\right]$ if and only if exists a real, positive number $M_f$, such that
\begin{equation}
	|\langle f | {\hat{T}_3^{(l)}} | g \rangle| \le M_f || g ||
	\label{cond_DT3dag}
\end{equation}
for any $g \in \cD(\hat{T}_3^{(l)})$.
If~(\ref{cond_DT3dag}) is true, then (by the Riesz–Fr\'echet theorem~\cite{RieszNagy:book}) exists $F \in L^{(2,l)}$, such that $\langle f | {\hat{T}_3^{(l)}} | g \rangle = \langle F | g \rangle$ and we define the adjoint operator by the relation $\left({\hat{T}_3^{(l)}}\right)^\dagger f = F$.
It is obvious that if $f \in \cD(\hat{T}_3^{(l)})$, then relation~(\ref{cond_DT3dag}) is satisfied and $F = {\hat{T}_3^{(l)}} f$.
This implies that $\cD(\hat{T}_3^{(l)}) \subseteq \cD\left[\left({\hat{T}_3^{(l)}}\right)^\dagger\right]$.
Therefore, to prove that ${\hat{T}_3^{(l)}}$ is self-adjoint (knowing already that it is hermitian), we have to show that also $\cD(\hat{T}_3^{(l)}) \supseteq \cD\left[\left({\hat{T}_3^{(l)}}\right)^\dagger\right]$.
For this, we calculate
\begin{eqnarray}
	\langle f | {\hat{T}_3^{(l)}} | g \rangle &\equiv& \int_0^L dl\, \rho f^*(l) \left( - \frac{i\hbar}{10 m_p} \left\{\left[ z\rho \hat{\brho} - \left( 2\rho^2 + z^2 \right) \hat{\bz} \right] \cdot \hat{\bl} \frac{\partial}{\partial l}
	+ \frac{T_l(l,q)}{2 \rho h_l} \right\}\right) g(l)
	\nonumber \\
	&=& - \frac{i\hbar}{10 m_p} \left[f^*(L) - f^*(0)\right] \left( \rho g \left\{\left[ z\rho \hat{\brho} - \left( 2\rho^2 + z^2 \right) \hat{\bz} \right] \cdot \hat{\bl} \right\}\right)_{l=0}
	+ \langle {\hat{T}_3^{(l)}} f | g \rangle
	\label{proof_SA}
\end{eqnarray}
where we used the properties of ${\hat{T}_3^{(l)}}$ derived in Section~\ref{subsec_quasi2D}.
The property~(\ref{cond_DT3dag}) is satisfied if in Eq.~(\ref{proof_SA}) $\hat{T}_3^{(l)} f \in L^{(2,l)}$ and $f^*(L) - f^*(0)$, that is, if and only if $f \in \cD(\hat{T}_3^{(l)})$.
This implies that $\cD(\hat{T}_3^{(l)}) \supseteq \cD\left[\left({\hat{T}_3^{(l)}}\right)^\dagger\right]$, which concludes the proof.

\section{The quantum particle on a torus with thin walls} \label{sec_eign_val_tor}

\begin{figure}[t]
	\centering
	\includegraphics[width=4cm]{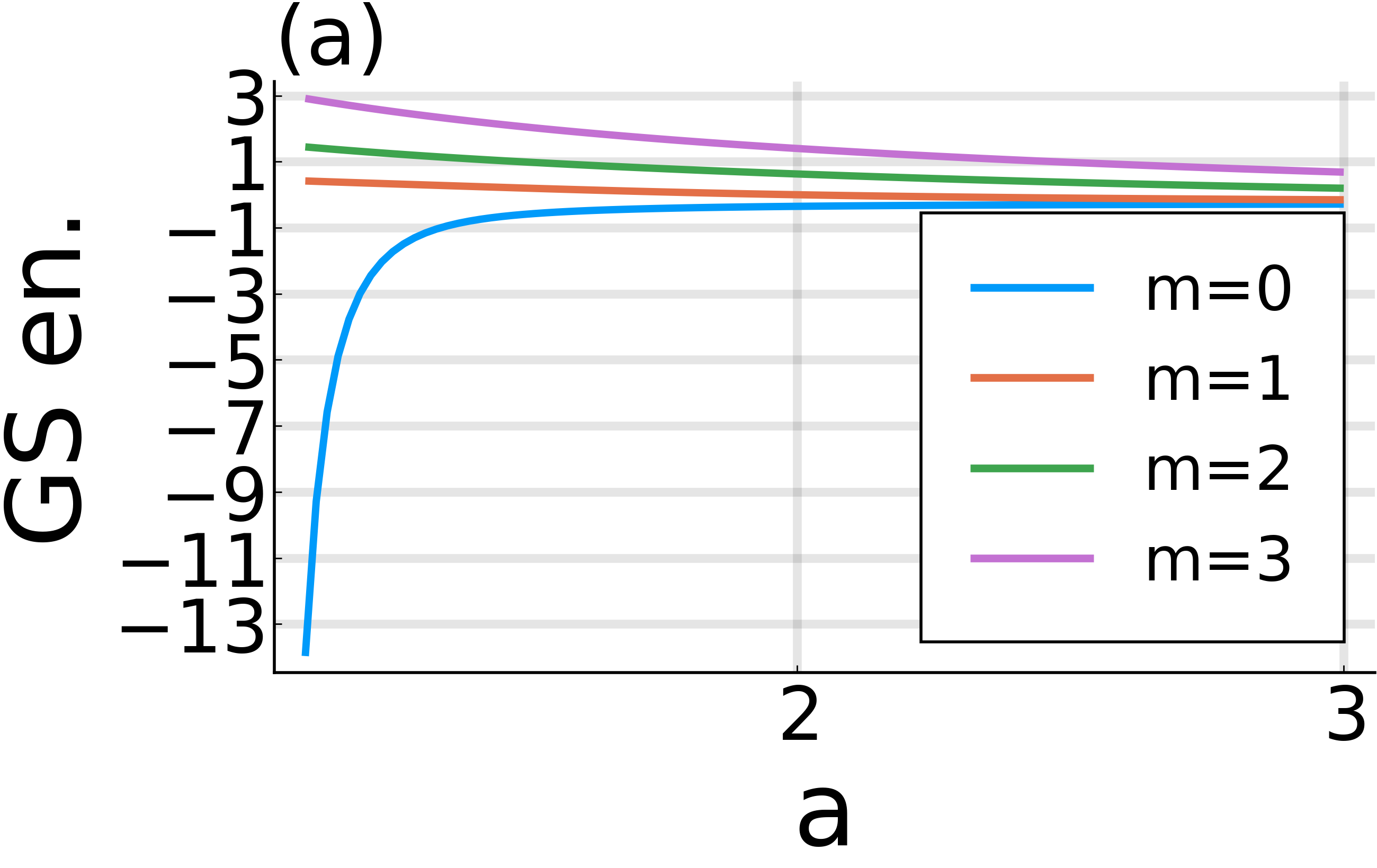}
	\hspace{3mm}
	\includegraphics[width=4cm]{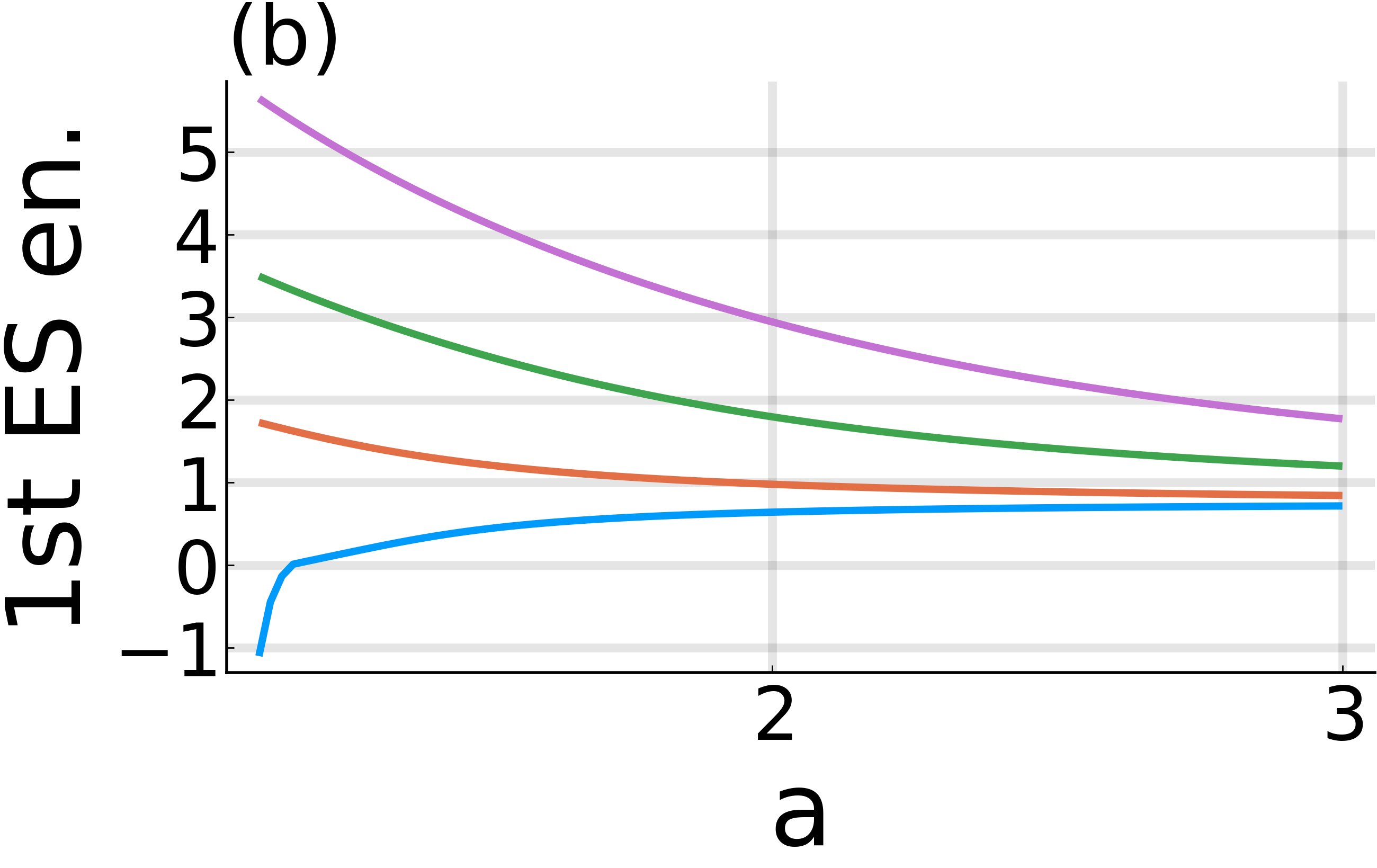}
	\hspace{3mm}
	\includegraphics[width=4cm]{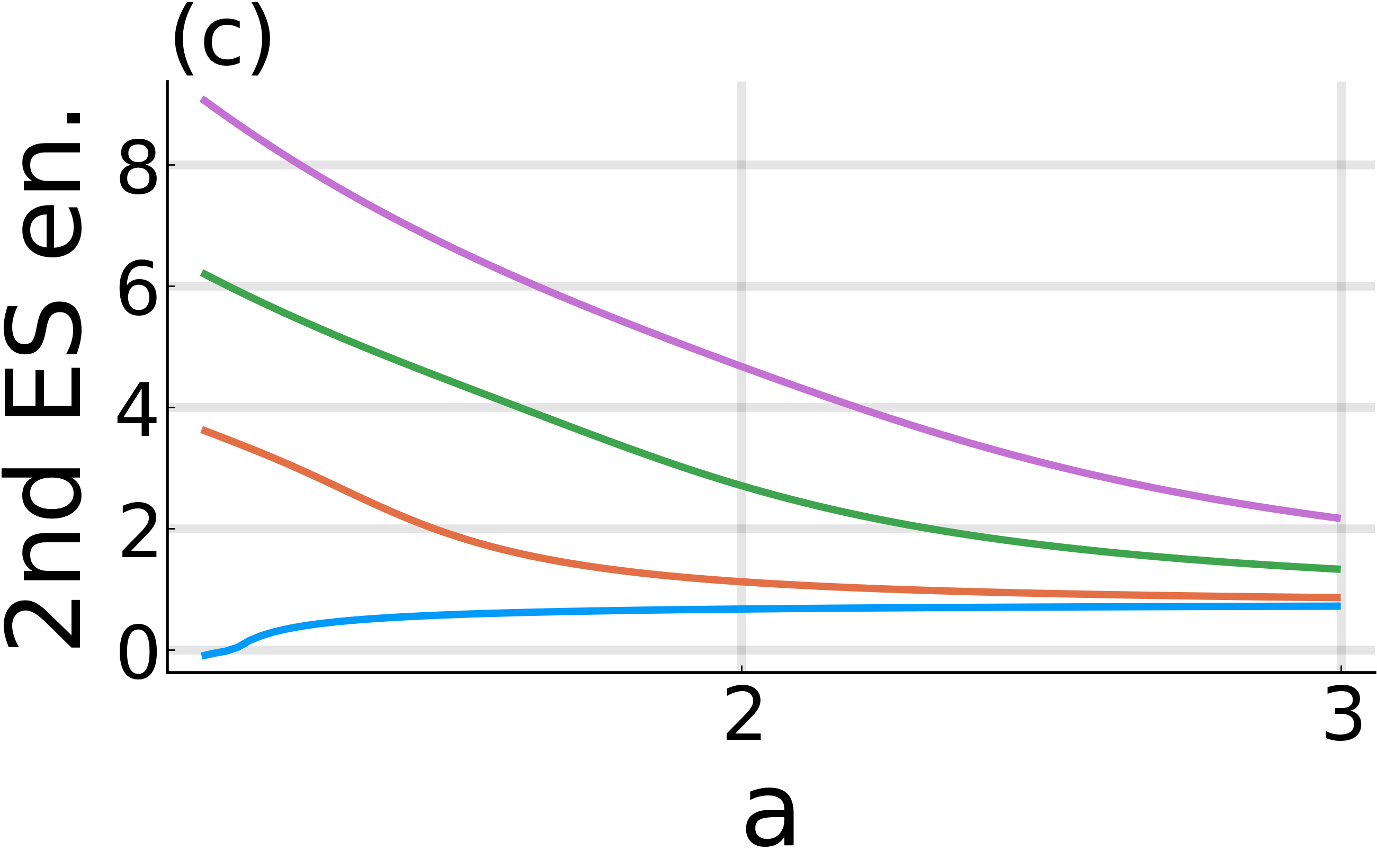}
	\hspace{3mm}
	\includegraphics[width=4cm]{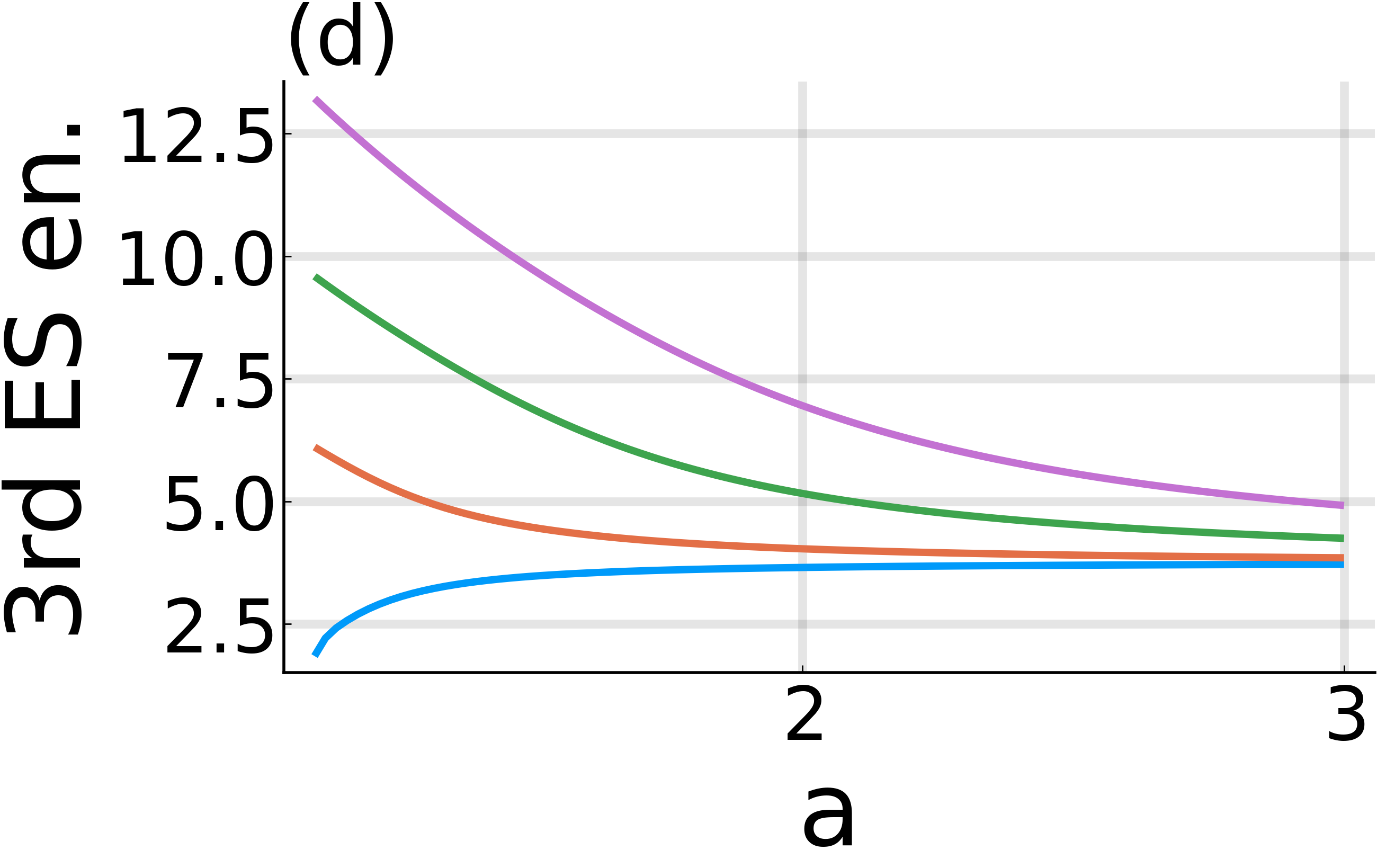}
	\caption{The energies of the ground state (a), first excited state (b), second excited state (c), and third excited state (d), as functions of $a$ and different values of $m$ (as shown in the legend).
	The legend is shown in (a) and is the same for all four figures.
	The energies are in units of $\hbar^2 a^2/(2m_pR^2)$ and include only the contribution of $\hat{\cH}^{(2D)}$ (without $\hat{\cH}^{(q)}$).
	}
	\label{fig_gs_1_2_3_en_lev}
\end{figure}

\begin{figure}[b]
	\centering
	\includegraphics[width=4cm]{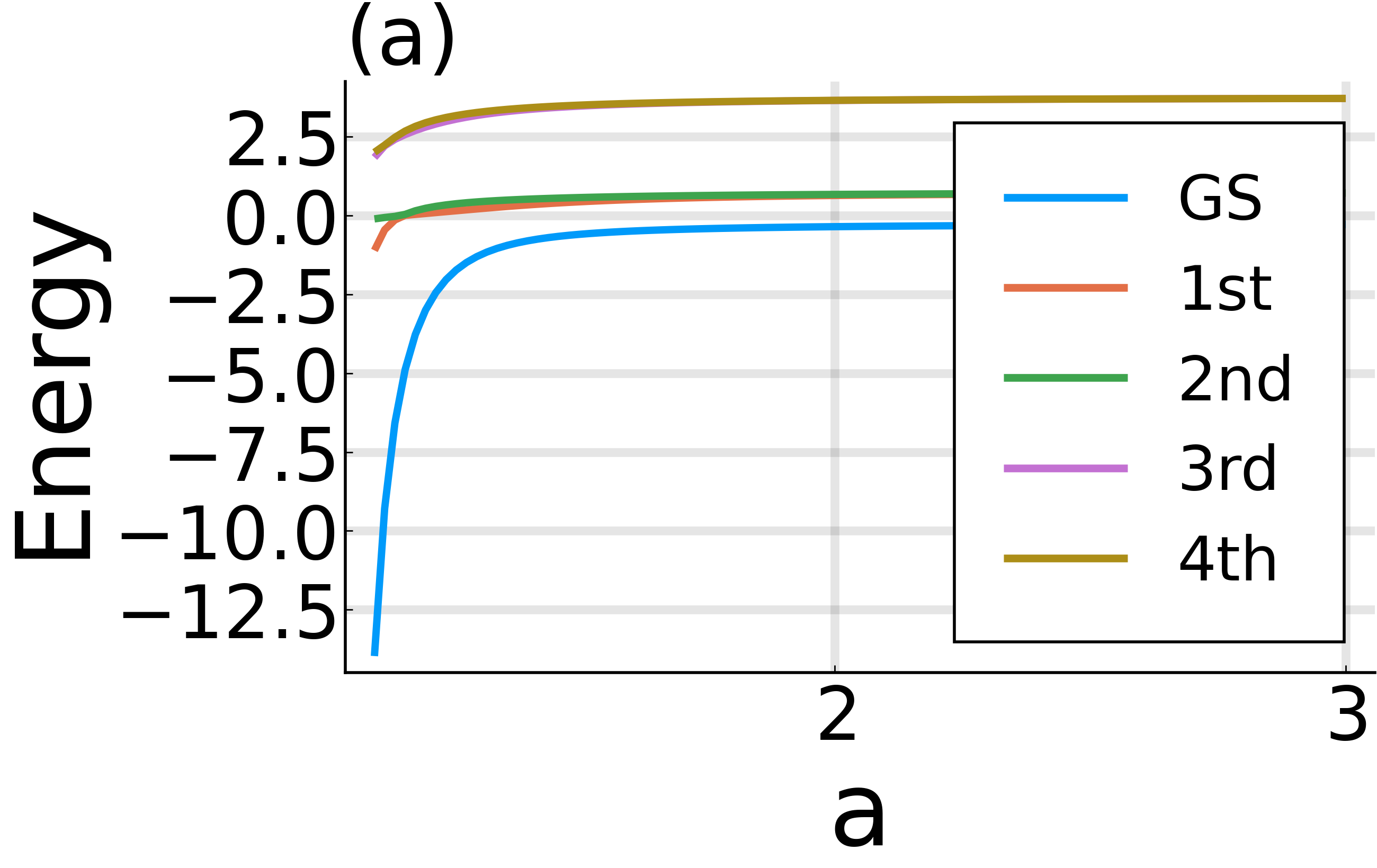}
	\includegraphics[width=4cm]{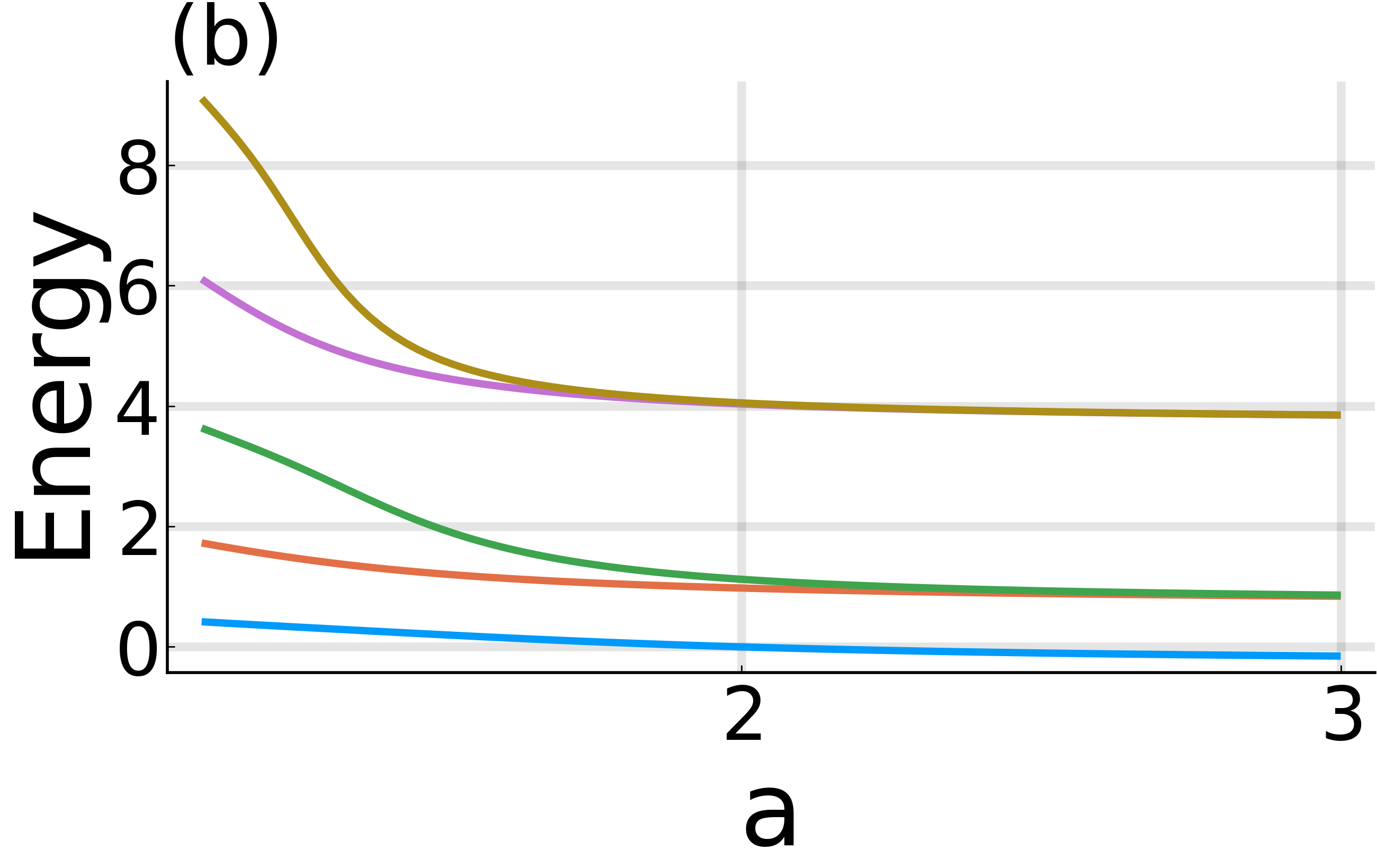}
	\includegraphics[width=4cm]{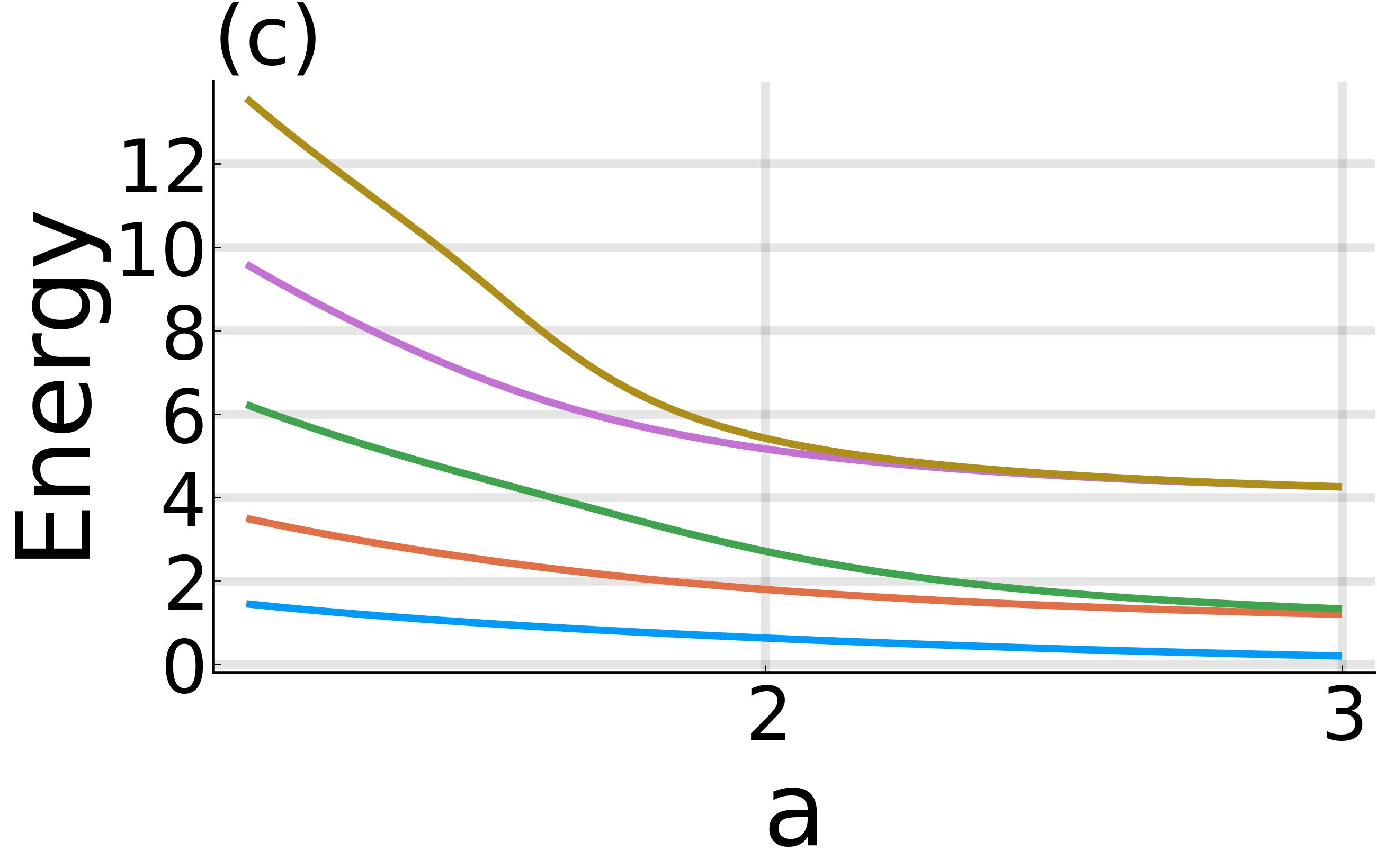}
	\includegraphics[width=4cm]{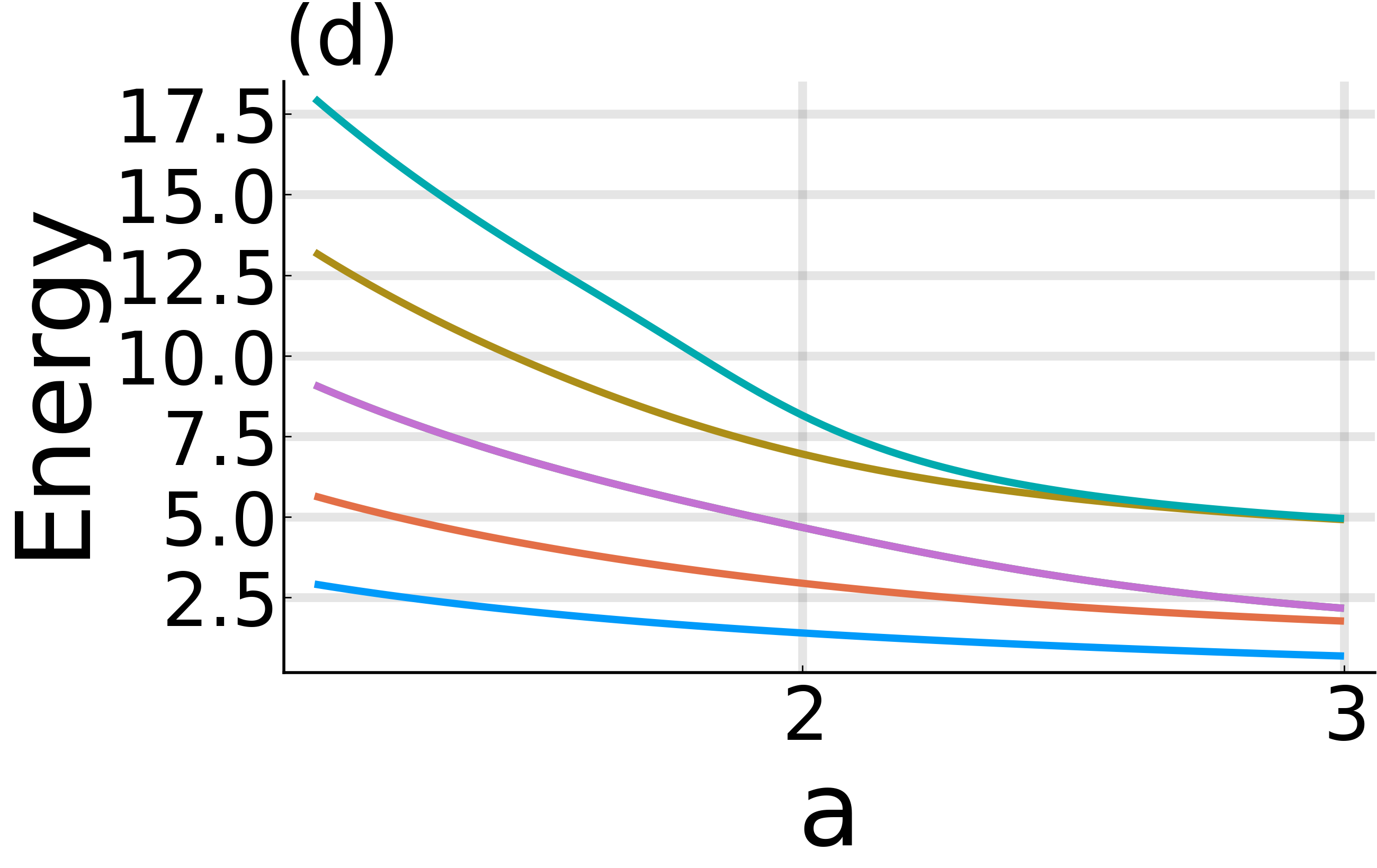}
	\caption{The energy levels vs. $a$, for different values of $m$: (a) $m=0$, (b) $m=1$, (c) $m=2$, (d) $m=3$.
	The legend, shown in (a), is the same for all figures.
	The energies are in units of $\hbar^2 a^2/(2m_pR^2)$ and include only the contribution of $\hat{\cH}^{(2D)}$ (without $\hat{\cH}^{(q)}$).
	}
	\label{fig_states_m0_m1_m2_m3}
\end{figure}

We apply the general formulae of Section~\ref{sec_CSCO} to calculate the toroidal dipole and energy eigenvalues on a torus with thin walls (quasi-2D system).
The torus has the major and minor radii $R$ and $r$, respectively, and its cross-section is presented in Fig.~\ref{fig_contour}~(b) and (c) (in Fig.~\ref{fig_contour}~c, the thickness of the wall is exaggerated for clarity).
We shall consider only tori with the ratio $a \equiv R/r >1$.
In this case, the coordinate $l$ is along the circle and the curvature radius $r(l)$ is independent of $l$.
We introduce the angle $\theta$, such that $l = r \theta$ (see Fig.~\ref{fig_contour}).
Then, $h_\theta = r+q \approx r$ and in the quasi-2D case the relevant operators are
\begin{subequations} \label{operators_torus}
\begin{eqnarray}
	\hat{p}^{(\bl)} &=& -i\hbar \left\{ \frac{\partial}{\partial l} - \frac{\sin\left(\frac{l}{r}\right)}{2\left[R+r\cos\left(\frac{l}{r}\right)\right]} \right\}
	\equiv -i\hbar \left\{ \frac{1}{r} \frac{\partial}{\partial \theta} - \frac{\sin\theta}{2 (R+r\cos\theta)} \right\}
	\label{pl_torus} \\
	%%%%%%%%%%%%%%%%%%%%%%%%%%%
	\hat{\cH}^{(l)} &=& \frac{-\hbar^2}{2m_p} \left( \frac{1}{R+r\cos\left(\frac{l}{r}\right)}
	\left\{ \left[R+r\cos\left(\frac{l}{r}\right)\right] \frac{\partial^2}{\partial l^2}
	- \sin\left(\frac{l}{r}\right) \frac{\partial}{\partial l} \right\}
	+ \frac{1}{4}\left[\frac{\cos\left(\frac{l}{r}\right)}{R+r\cos\left(\frac{l}{r}\right)} - \frac{1}{r}\right]^2 \right) \nonumber \\
	&\equiv& \frac{-\hbar^2}{2m_p} \left( \frac{1}{R+r\cos\theta}
	\left\{ \frac{R+r\cos\theta}{r^2} \frac{\partial^2}{\partial \theta^2}
	- \frac{\sin\theta}{r} \frac{\partial}{\partial\theta} \right\}
	+ \frac{1}{4}\left[\frac{\cos\theta}{R+r\cos\theta} - \frac{1}{r}\right]^2 \right)
%	\nonumber \\
%	%
%	&\equiv&\frac{-\hbar^2}{2m_p} \left[ \frac{1}{\rho}
%	\left( \frac{\rho}{r^2} \frac{\partial^2}{\partial\theta^2}
%	- \frac{\sin\theta}{r} \frac{\partial}{\partial\theta} \right)
%	+ \frac{1}{4}\left(\frac{\cos\theta}{\rho} - \frac{1}{r}\right)^2 \right] ,
	\label{Hl_torus} \\
	%%%%%%%%%%%%%%%%%%%%%%%%
	\hat{\cH}^{(\phi)} &=& \frac{-\hbar^2}{2m_p} \frac{1}{\left[R+r\cos\left(\frac{l}{r}\right)\right]^2} \frac{\partial^2}{\partial \phi^2}
	\equiv \frac{-\hbar^2}{2m_p} \frac{1}{\left[R+r\cos\theta\right]^2} \frac{\partial^2}{\partial \phi^2} ,
	\label{Hphi_torus} \\
	%%%%%%%%%%%%%%%%%%%%%%%%
	\hat{T}^{(l)}_3 &=& - \frac{i\hbar}{10 m_p} \left( - z\left[R+r\cos\left(\frac{l}{r}\right)\right] \sin\left(\frac{l}{r}\right) - \left\{ 2\left[R+r\cos\left(\frac{l}{r}\right)\right]^2 + z^2 \right\} \cos\left(\frac{l}{r}\right) \right) \frac{\partial}{\partial l}
	-
	\frac{i\hbar}{20 m_p \rho} T_l(l,0)
	\nonumber \\
	&=& - \frac{i\hbar}{10 m_p} \frac{ - r\sin\theta (R+r\cos\theta) \sin\theta - \left[ 2 (R+r\cos\theta)^2 + r^2 \sin^2\theta \right] \cos\theta }{r} \frac{\partial}{\partial \theta}
	-
	\frac{i\hbar}{20 m_p \rho} T_l(\theta,0) ,
%	\nonumber \\
%	%
%	&=& - \frac{i\hbar}{10 m_p} \frac{- z\rho \sin\theta - \left( 2\rho^2 + z^2 \right) \cos\theta}{r} \frac{\partial}{\partial\theta}
%	-
%	\frac{i\hbar}{20 m_p \rho r} T_l(\theta,0)
	\label{T3_torus1} \\
	%%%%%%%%%%%%%%%%%%%%%%
	T_l(l) &=& 2z \left[ R+r \cos\left(\frac{l}{r}\right) \right] \left[ \sin^2\left(\frac{l}{r} \right) - \cos^2\left(\frac{l}{r}\right) \right]
	+ \left\{5\left[R+r\cos\left(\frac{l}{r}\right)\right]^2 + z^2\right\} \sin\left(\frac{l}{r}\right) \cos\left(\frac{l}{r}\right)
	\nonumber \\
	&& + \frac{1}{r} \left[ R+r \cos\left(\frac{l}{r}\right) \right] \left(
	- z \left[ R+r \cos\left(\frac{l}{r}\right) \right] \cos\left(\frac{l}{r}\right) + \left\{ 2\left[ R+r \cos\left(\frac{l}{r}\right) \right]^2 + z^2 \right\} \cos\left(\frac{l}{r}\right) \right)
	\nonumber \\
	&\equiv& - 2 r\sin\theta (R+r \cos\theta) \cos(2\theta)
	+ \left[ 5 (R+r\cos\theta)^2 + r^2 \sin^2\theta \right] \frac{\sin(2\theta)}{2}
	+ \frac{R+r \cos\theta}{r}
	\nonumber \\
	&& \times \left\{
	( 2 R + r \cos\theta ) (R+r \cos\theta) + r^2 \sin^2\theta \right\} \sin\theta .
	\label{T3_torus2}
\end{eqnarray}
\end{subequations}
The momenta eigenfunctions on the torus are
\begin{equation} \label{q_cond_tor}
\cF^{(\Lambda)}_{k,m} (l,\phi) \equiv \frac{1}{\sqrt{2\pi L}}\frac{e^{i (l k + m\phi)}}{\sqrt{R+r \cos\left(\frac{l}{r}\right)}}
\quad {\rm or} \quad
\cF^{(\Lambda)}_{n,m} \equiv \frac{1}{2\pi}\frac{e^{i (n \theta + m\phi)}}{\sqrt{R+r \cos\theta}} ,
\quad {\rm where} \quad
k = \frac{2\pi n}{L}, \quad n,m = {\rm integers} .
\end{equation}
This simplifies considerably the calculations and allows us to obtain analytical expressions for matrix elements of $\hat{T}_3$ and $\hat{\cH}^{(2D)}$.
For the Hamiltonian we obtain
\begin{equation}
	\langle \cF^{(\Lambda)}_{n_1,m_1} | \hat{\cH}^{(2D)} | \cF^{(\Lambda)}_{n_2,m_2} \rangle
	= \frac{\hbar^2}{2m_p} \frac{a^2}{R^2}
	\left[\left(n_1^{2}-\frac{1}{4}\right) \delta_{n_1n_2}
	+\frac{\left({|\delta n|} \sqrt{a^{2}-1}+a \right) \left(\sqrt{a^{2}-1}-a \right)^{{| \delta n |}}}{\left(a^{2}-1\right)^{\frac{3}{2}}} \left(m_1^{2}-\frac{1}{4}\right)
	\right]  \delta_{m_1m_2}
	\label{H_elms_matr}
\end{equation}
where $\delta n \equiv n_1-n_2$ ($a \equiv R/r$, as defined before).
Similarly, for the toroidal dipole we have
%
%\begin{subequations} \label{T3_elms_matr}
\begin{eqnarray}
	\langle \cF^{(\Lambda)}_{n_1,m_1} | \hat{T}_3^{(l)} | \cF^{(\Lambda)}_{n_2,m_2} \rangle
%	&=&  - \frac{\hbar}{10 m_p} r \left\{ n_2 \left[\frac{5 a \delta_{n_1n_2}}{2}+\left(a^{2}+1\right) \left(\delta_{n_1,n_2+1} +\delta_{n_1,n_2-1} \right)+\frac{3 a \left(\delta_{n_1,n_2+2} +\delta_{n_1,n_2-2} \right)}{4}\right]
%	\right. \nonumber \\
%	%
%	&& \left. + \left[-\frac{\left(a^{2}+1\right) \left(\delta_{n_2,n_1+1} - \delta_{n_2,n_1-1} \right)}{2}-\frac{3 a \left(\delta_{n_2,n_1+2} -\delta_{n_2,n_1-2} \right)}{4}\right]
%	\right\} \delta_{m_1m_2}
%	\nonumber \\
%	%
	&=&  - \frac{\hbar}{10 m_p} \frac{a}{R} \frac{n_1+n_2}{2} \left\{ \frac{5 a}{2} \delta_{n_1n_2}
	+ \left(a^{2}+1\right) \left(\delta_{n_1,n_2+1} + \delta_{n_1,n_2-1} \right)
	\right. \nonumber \\
	&& \left. + \frac{3 a }{4} \left(\delta_{n_1,n_2+2} +\delta_{n_1,n_2-2} \right)
	\right\} \delta_{m_1m_2} .
	\label{T3_elms_matr2}
\end{eqnarray}
%\end{subequations}
%
From Eq.~(\ref{H_elms_matr}) we observe that the diagonal elements are
\begin{subequations} \label{H_diag_nondiag}
\begin{eqnarray}
	\langle \cF^{(\Lambda)}_{n,m} | \hat{\cH}^{(2D)} | \cF^{(\Lambda)}_{n,m} \rangle
	&=& \frac{\hbar^2}{2m_p} \frac{a^2}{R^2}
	\left[\left(n^{2}-\frac{1}{4}\right)
	+\frac{a}{\left(a^{2}-1\right)^{\frac{3}{2}}} \left(m^{2}-\frac{1}{4}\right)
	\right]
	\label{H_diag}
\end{eqnarray}
whereas the non-diagonal elements, $n_1 \ne n_2$,
\begin{equation}
	\langle \cF^{(\Lambda)}_{n_1,m_1} | \hat{\cH}^{(2D)} | \cF^{(\Lambda)}_{n_2,m_2} \rangle
	= \frac{\hbar^2}{2m_p} \frac{a^2}{R^2}
	\frac{\left({|\delta n|} \sqrt{a^{2}-1}+a \right) \left(\sqrt{a^{2}-1}-a \right)^{{| \delta n |}}}{\left(a^{2}-1\right)^{\frac{3}{2}}} \left(m_1^{2}-\frac{1}{4}\right)
	\delta_{m_1m_2}
\end{equation}
\end{subequations}
decrease (roughly) in geometrical progression with $|\delta n|$, because of the factor $\left(\sqrt{a^{2}-1}-a \right)^{{| \delta n |}}$, where $|\sqrt{a^{2}-1}-a| < 1$ and $a>1$.
Because of this fast decrease of the non-diagonal elements, the secular equation can be solved easily and the solution converges fast with the dimension of the matrix.
For example, the maximum absolute value of the difference between the eigenvalues calculated with matrices of dimensions $501 \times 501$ (shown in Figs.~\ref{fig_gs_1_2_3_en_lev} and \ref{fig_states_m0_m1_m2_m3}) and with matrices of dimensions $20001 \times 20001$ is $5\times 10^{-6}$.

\begin{figure}[t]
	\centering
	\includegraphics[width=7cm]{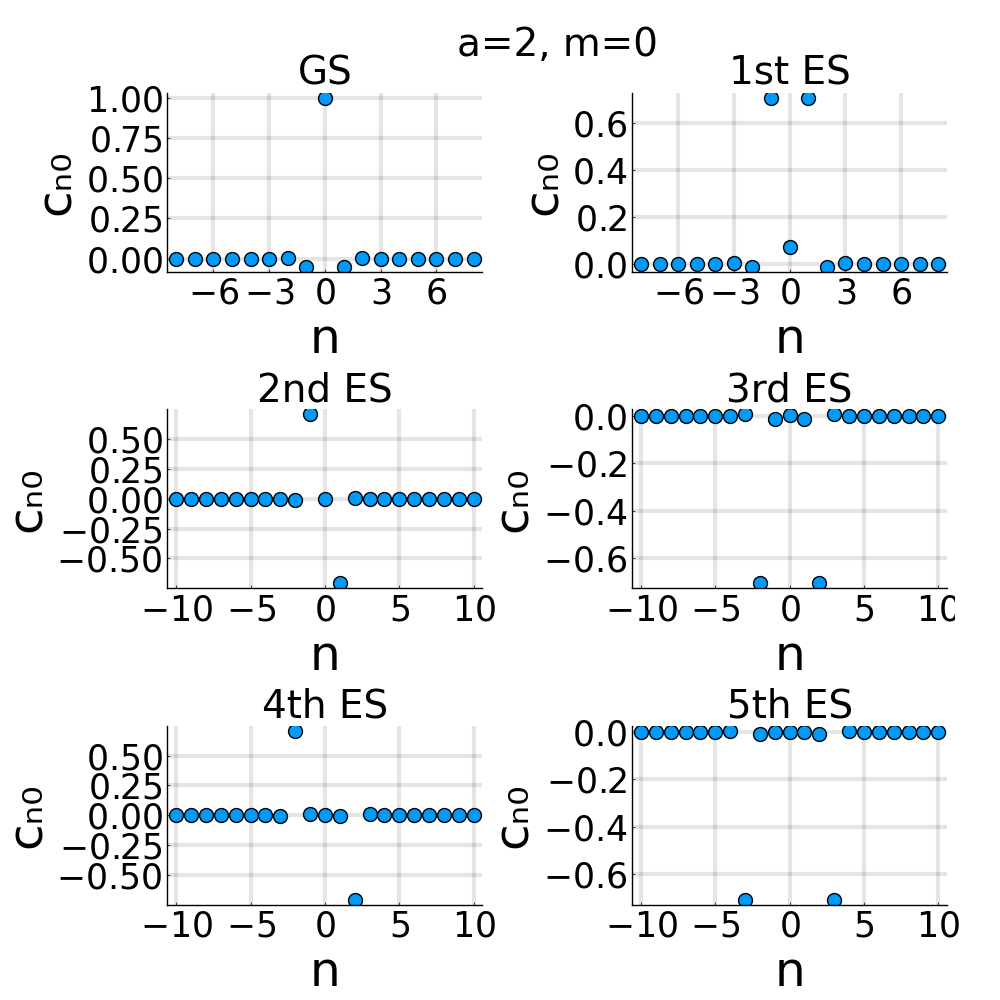}
	\hspace{1cm}
	\includegraphics[width=7cm]{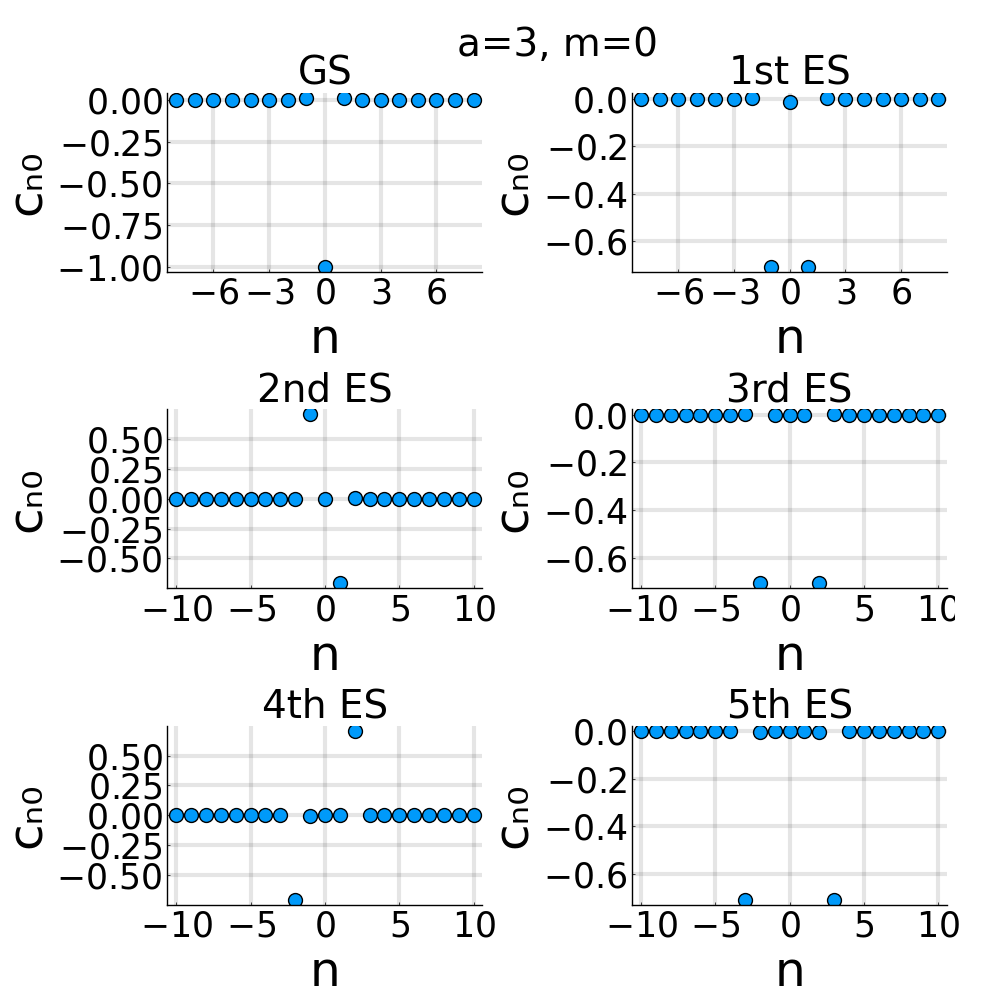}
	\caption{The coefficients of the expansion of the energy eigenvectors in the basis formed by the functions $\cF^{(\Lambda)}_{n,m} (\theta, \phi)$, for $m=0$, $a = 2$ (left) and $a = 3$ (right).}
	\label{fig_vectors1-5_m0}
\end{figure}

In Fig.~\ref{fig_gs_1_2_3_en_lev} we show the energies of the ground state, $1^{\rm st}$, $2^{\rm nd}$, and $3^{\rm rd}$ excited states, as functions of $a$ and for the angular momentum quantum number $m = 0,1,2,3$.
We observe that the lines converge, as $a$ increases. This is due to the decrease with $a$ of the influence of $m$ on the matrix elements~(\ref{H_elms_matr}).
For comparison, in Fig.~\ref{fig_states_m0_m1_m2_m3} we show the first five energy levels as functions of $a$, for different values of $m = 0, 1, 2, 3$.
We notice that for any value of $m$, the excited energy levels converge two-by-two as $a$ increases.
We can better understand this result if we think of the expansion of the energy eigenstates $\Psi^{\cH^{(l)}}$ in the basis formed by the set~(\ref{q_cond_tor}):
\begin{equation} \label{exp_PsiH1}
	\Psi^{\cH^{(2D)}}_{h,m,a} (\theta, \phi) \equiv \sum_{n=-\infty}^{\infty} C_{n,m}^{(h,a)} \cF^{(\Lambda)}_{n,m} (\theta, \phi) ,
\end{equation}
where $h$ represents the energy level and $m$ is the orbital quantum number.
In Fig.~\ref{fig_vectors1-5_m0} we plot the relevant coefficients $C_{n,m}^{(h)}$ for $h = 0,\ldots,5$, $a=2$ (left panel) and $a=3$ (right panel).
We see that for the ground state ($h=0$) only the coefficient $C_{0,m}^{(0)}$ is significantly different from zero, whereas for the excited states the eigenvectors are roughly symmetric or antisymmetric superpositions of the functions $\cF^{(\Lambda)}_{n,m} (\theta, \phi)$.
We may approximately write
\begin{eqnarray}
	&& \Psi^{\cH^{(2D)}}_{0,0,a} (\theta, \phi) \approx \cF^{(\Lambda)}_{0,0} (\theta, \phi) ,
	\quad
	\Psi^{\cH^{(2D)}}_{1,0,a} (\theta, \phi) \approx \dfrac{1}{\sqrt{2}} \left[ \cF^{(\Lambda)}_{-1,0} (\theta, \phi) + \cF^{(\Lambda)}_{1,0} (\theta, \phi) \right] ,
	\nonumber \\
	&& \Psi^{\cH^{(2D)}}_{2,0,a} (\theta, \phi) \approx \dfrac{1}{\sqrt{2}} \left[ \cF^{(\Lambda)}_{-1,0} (\theta, \phi) - \cF^{(\Lambda)}_{1,0} (\theta, \phi) \right] ,
	\quad
	\Psi^{\cH^{(2D)}}_{3,0,a} (\theta, \phi) \approx \dfrac{1}{\sqrt{2}} \left[ \cF^{(\Lambda)}_{-2,0} (\theta, \phi) + \cF^{(\Lambda)}_{2,0} (\theta, \phi) \right] ,
	\nonumber \\
	&& \Psi^{\cH^{(2D)}}_{4,0,a} (\theta, \phi) \approx \dfrac{1}{\sqrt{2}} \left[ \cF^{(\Lambda)}_{-2,0} (\theta, \phi) - \cF^{(\Lambda)}_{2,0} (\theta, \phi) \right] ,
	\quad
	\Psi^{\cH^{(2D)}}_{5,0,a} (\theta, \phi) \approx \dfrac{1}{\sqrt{2}} \left[ \cF^{(\Lambda)}_{-3,0} (\theta, \phi) + \cF^{(\Lambda)}_{3,0} (\theta, \phi) \right] ,
	\label{approx_C_m0_0_5}
\end{eqnarray}
for both, $a=2$ and $a=3$ (the overall sign is not relevant).
Equations~(\ref{approx_C_m0_0_5}) explain qualitatively the (approximative) degeneracy of the excited energy levels.

\begin{figure}[t]
	\centering
	\includegraphics[width=7cm]{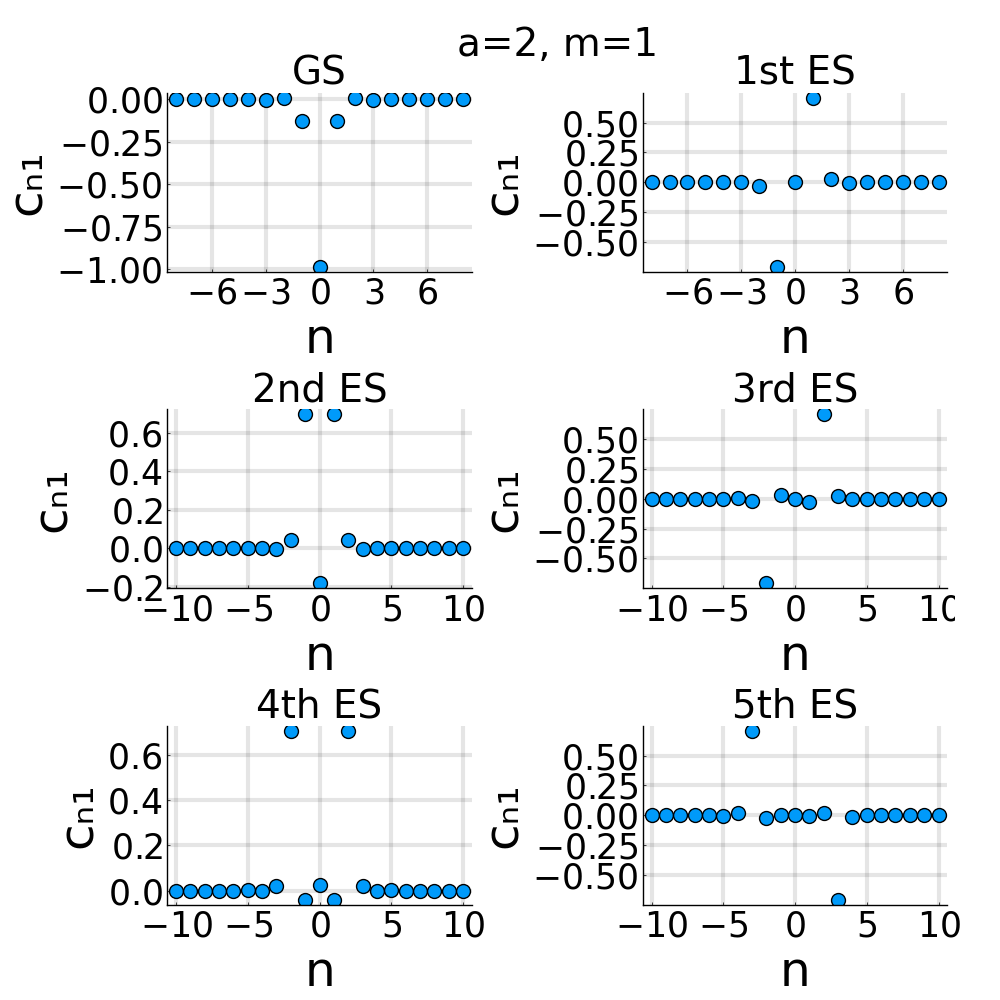}
	\hspace{1cm}
	\includegraphics[width=7cm]{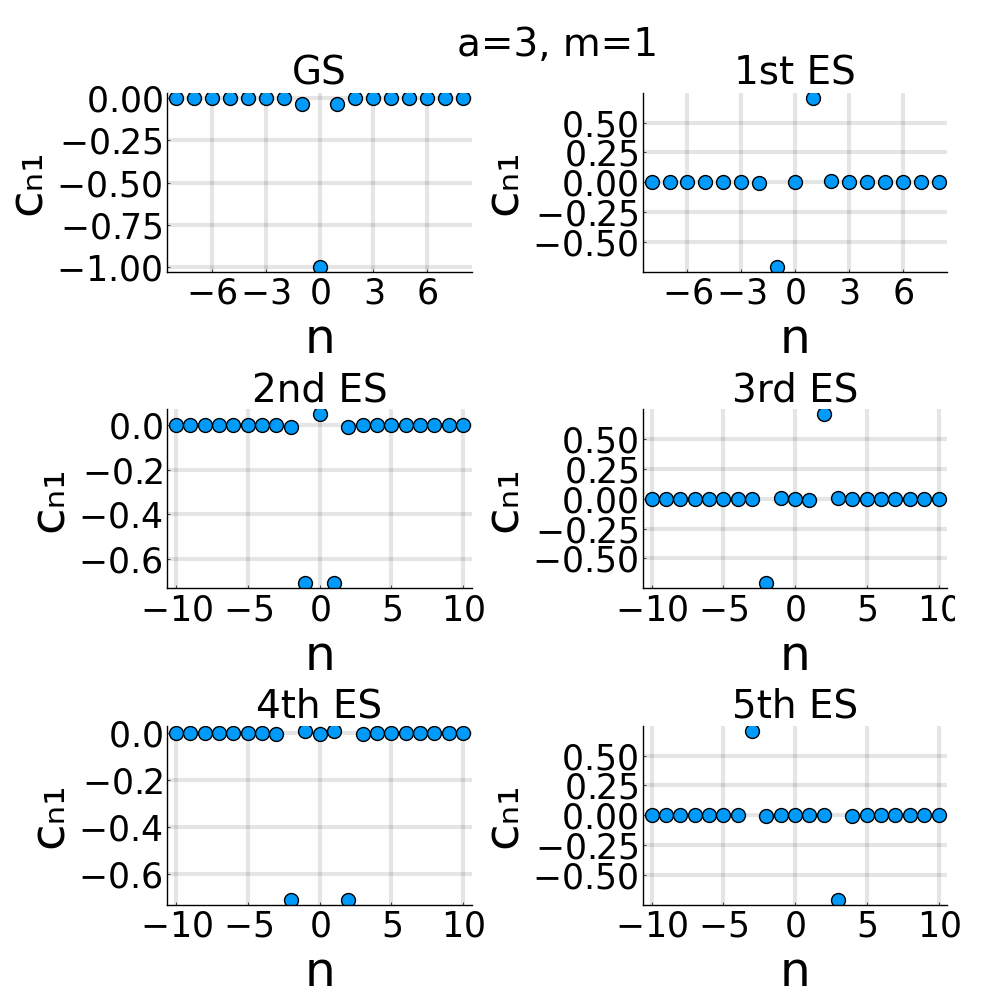}
	\caption{The coefficients of the expansion of the energy eigenvectors in the basis formed by the functions $\cF^{(\Lambda)}_{n,m} (\theta, \phi)$, for $m=1$, $a = 2$ (left) and $a = 3$ (right).}
	\label{fig_vectors1-5_m1}
\end{figure}

A similar situation appears also for the higher values of $m$, as it is exemplified in Fig.~\ref{fig_vectors1-5_m1} for $m=1$.
Nevertheless, the quasi-degeneracy appears at higher and higher values of $a$, as $m$ increases, as we can see in Fig.~\ref{fig_states_m0_m1_m2_m3}.
Furthermore, we observe that an energy gap is formed for $m=0$, between the ground state and the first excited state, as $a$ decreases towards 1.

\begin{figure}[b]
	\centering
	\includegraphics[width=7cm]{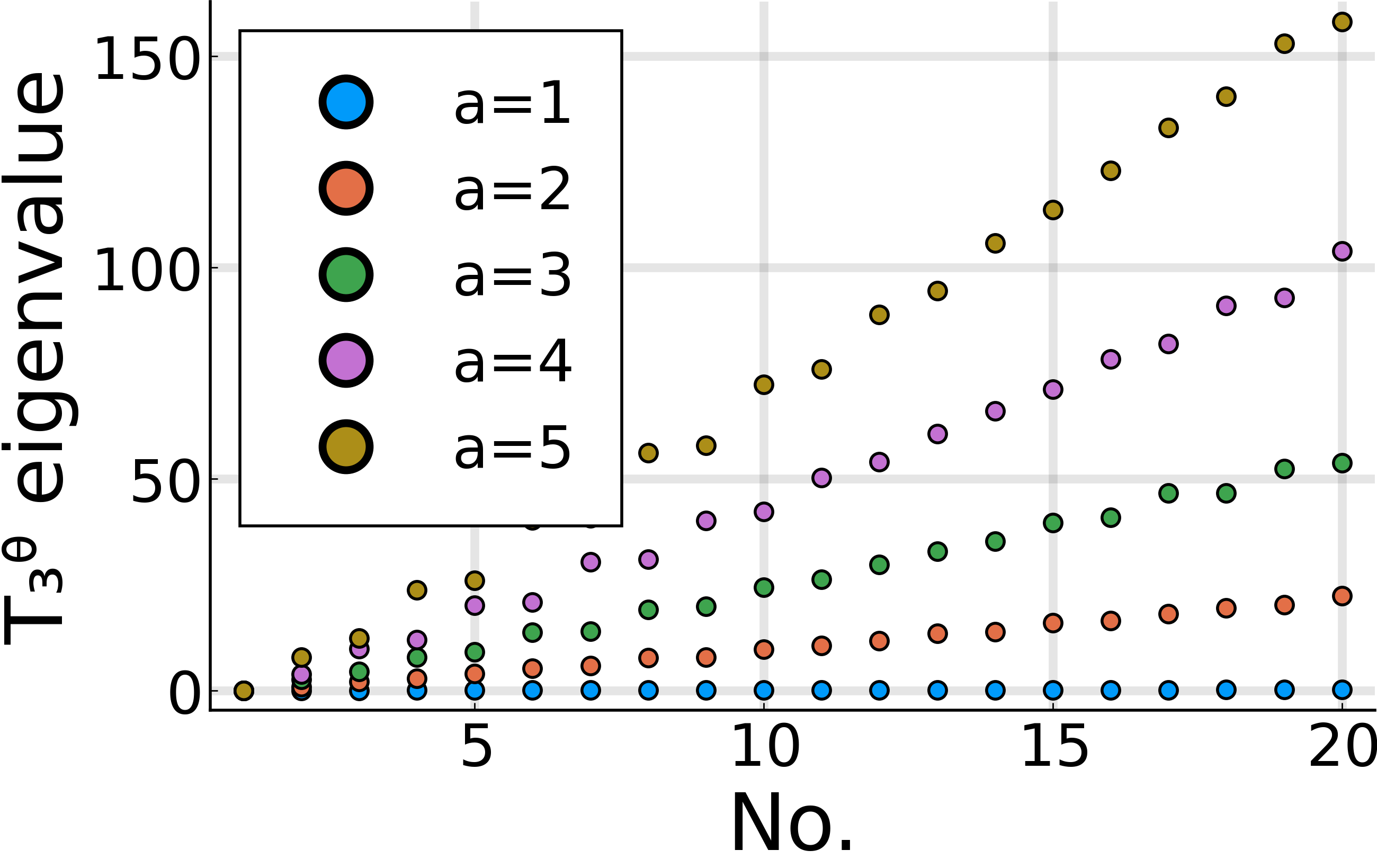}
	\caption{The eigenvalues of the toroidal dipole operator for four values of $a$.}
	\label{fig_tor10001}
\end{figure}

In Fig.~\ref{fig_tor10001} we plot the eigenvalues of the toroidal dipole operator for four different values of $a$.
In this case, the numerical calculations converge much slower than in the case of the Hamiltonian.
The results in Fig.~\ref{fig_tor10001} are obtained by diagonalizing matrices of dimension $10001 \times 10001$.
% and some structure due to the finite size of the matrix may be observed at higher eigenvalues.
We observe that the eigenvalues of $\hat{T}_3^{(l)}$ are almost linearly distributed and are separated by distances which increase with $a$.

\begin{figure}[t]
	\centering
	\includegraphics[width=4cm]{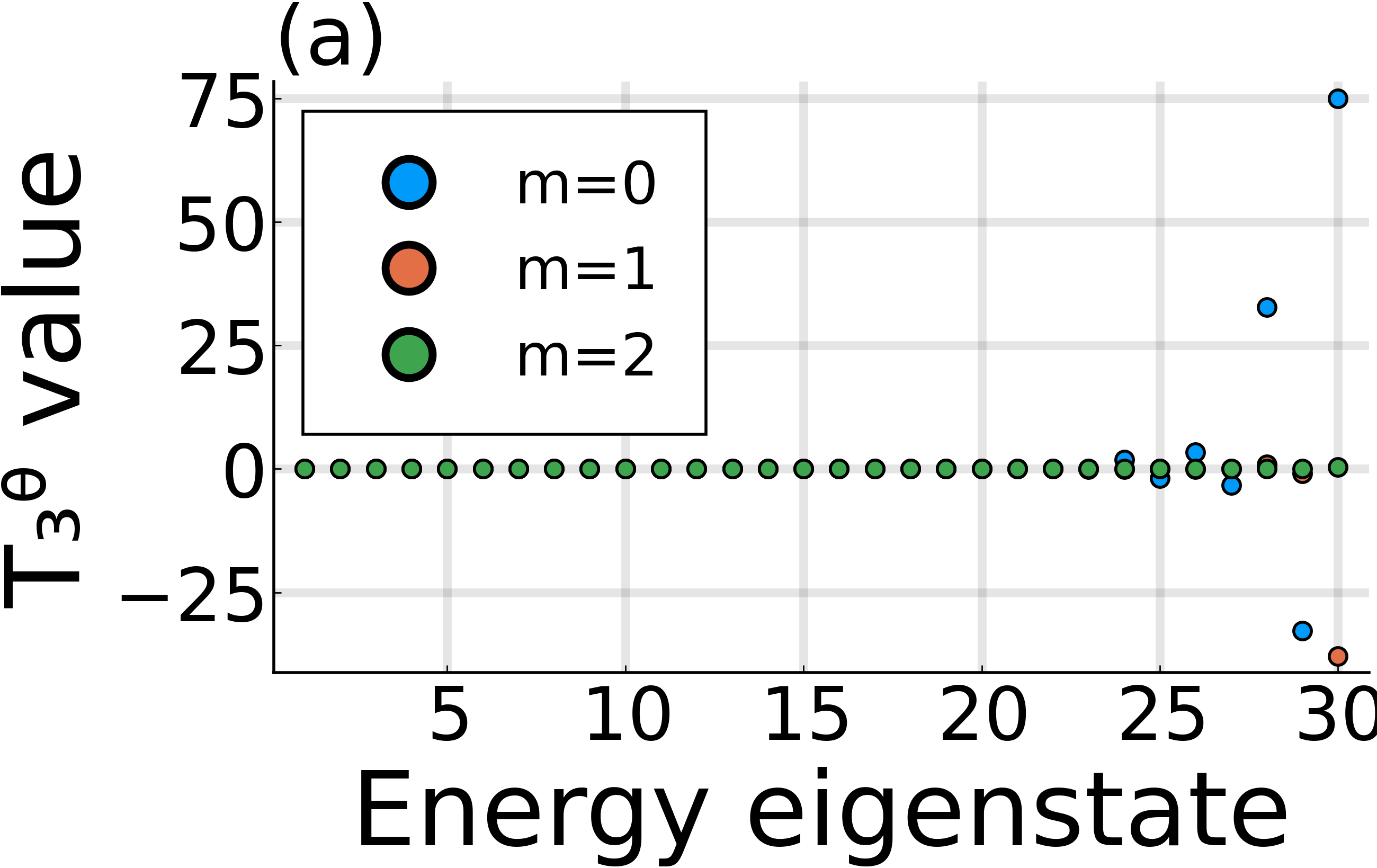}
	\includegraphics[width=4cm]{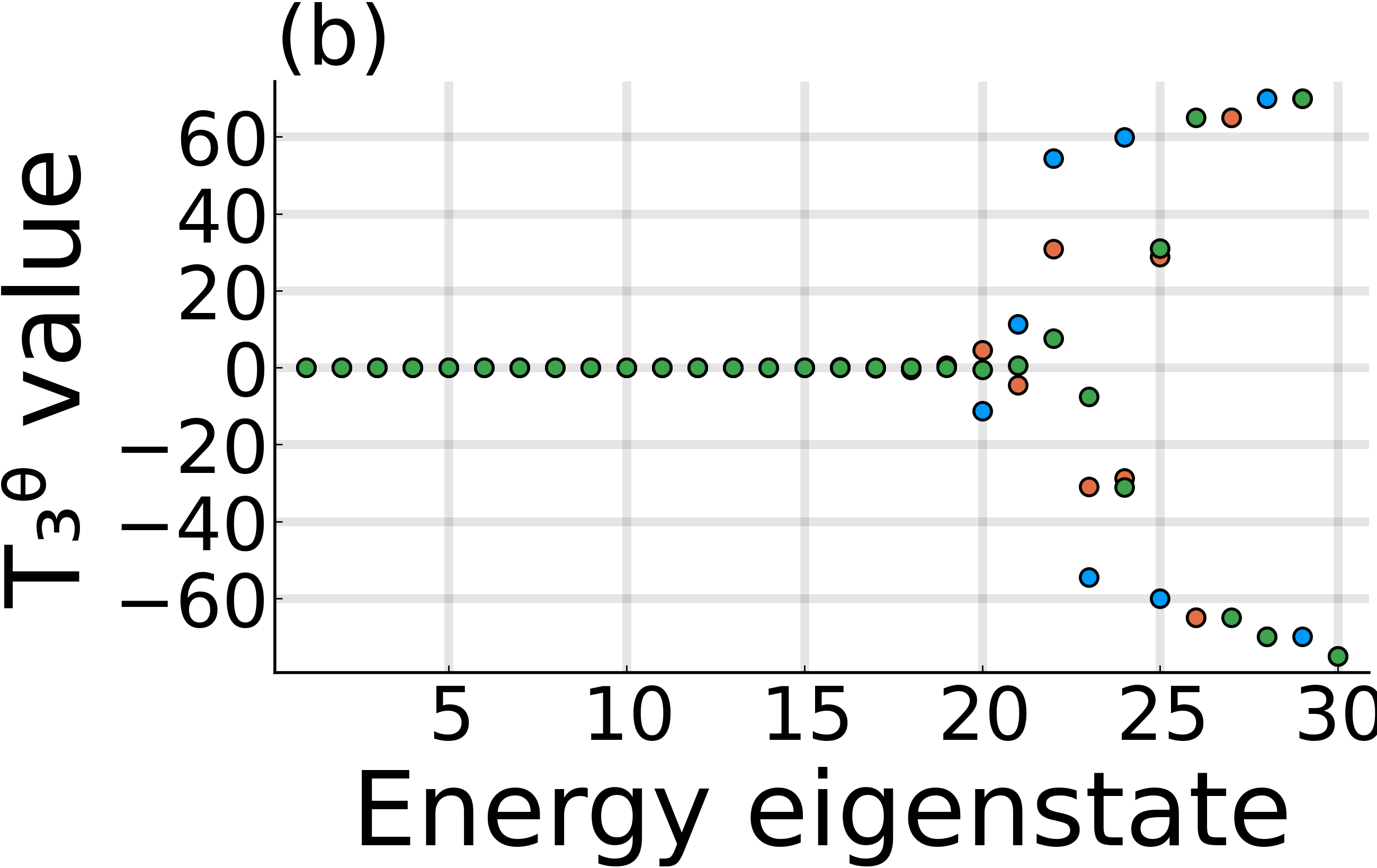}
	\includegraphics[width=4cm]{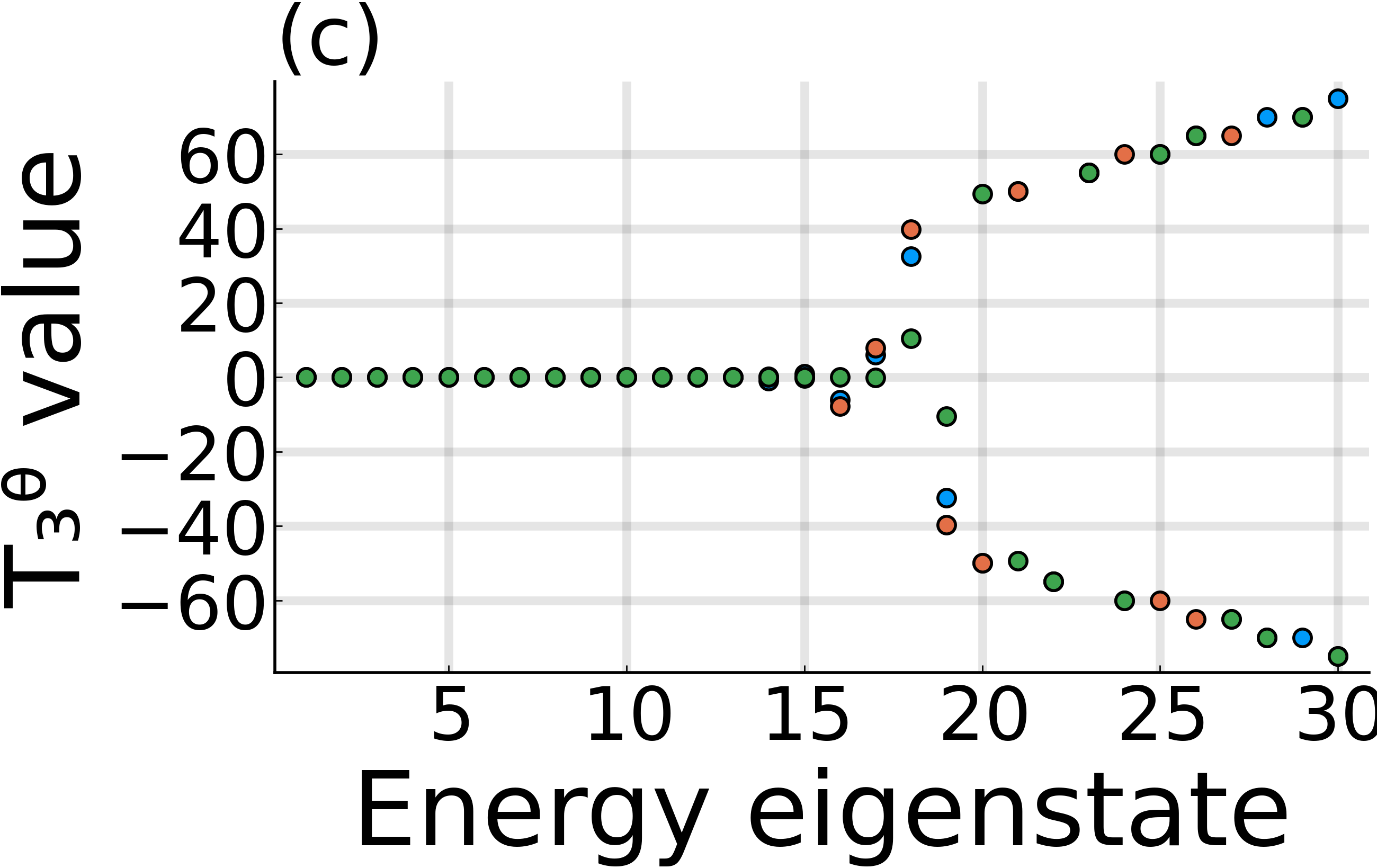}
	\includegraphics[width=4cm]{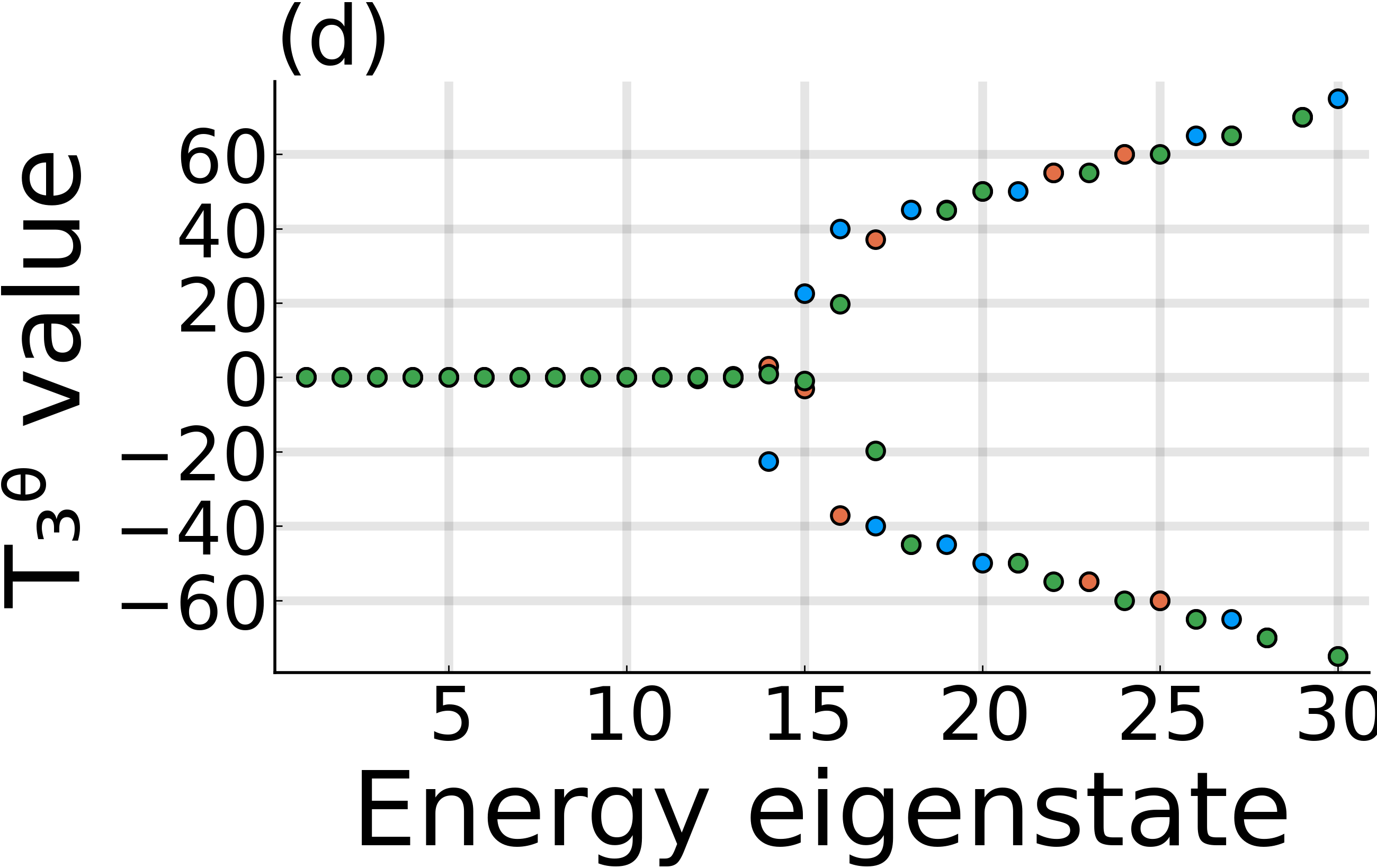}
	\caption{The expectation values of the toroidal dipole operator on the eigenstates of the Hamiltonian, for $a=1.5, 2, 2.5, 3$.}
	\label{fig_Tor_exps}
\end{figure}

It is interesting to calculate the expectation values of $\hat{T}_3^{(l)}$ on the Hamiltonian eigenstates.
We plot these in Fig.~\ref{fig_Tor_exps}, separately, for different values of $a$ and observe that up to a certain energy level the expectation values are almost zero.
They start to differ from zero at energy levels that depend on $a$ and have a weak dependence on $m$.
The low energy behavior may be understood based on Figs.~\ref{fig_vectors1-5_m0} and \ref{fig_vectors1-5_m1} and the approximations~(\ref{approx_C_m0_0_5}).
The coefficients of the expansion of the Hamiltonian eigenstates appear in pairs of the form $C_{n,m}^{(h,a)}$ and $C_{-n,m}^{(h,a)}$ and since the matrix elements~(\ref{T3_elms_matr2}) have the property
\begin{equation}
	\langle \cF^{(\Lambda)}_{n,m} | \hat{T}_3^{(l)} | \cF^{(\Lambda)}_{n,m} \rangle
	= - \langle \cF^{(\Lambda)}_{-n,m} | \hat{T}_3^{(l)} | \cF^{(\Lambda)}_{-n,m} \rangle
	\quad {\rm and} \quad
	\langle \cF^{(\Lambda)}_{n,m} | \hat{T}_3^{(l)} | \cF^{(\Lambda)}_{-n,m} \rangle = 0 ,
	\label{symmetry_T3}
\end{equation}
then they add to approximately zero in the expectation value.

\begin{figure}[b]
	\centering
	\includegraphics[width=7cm]{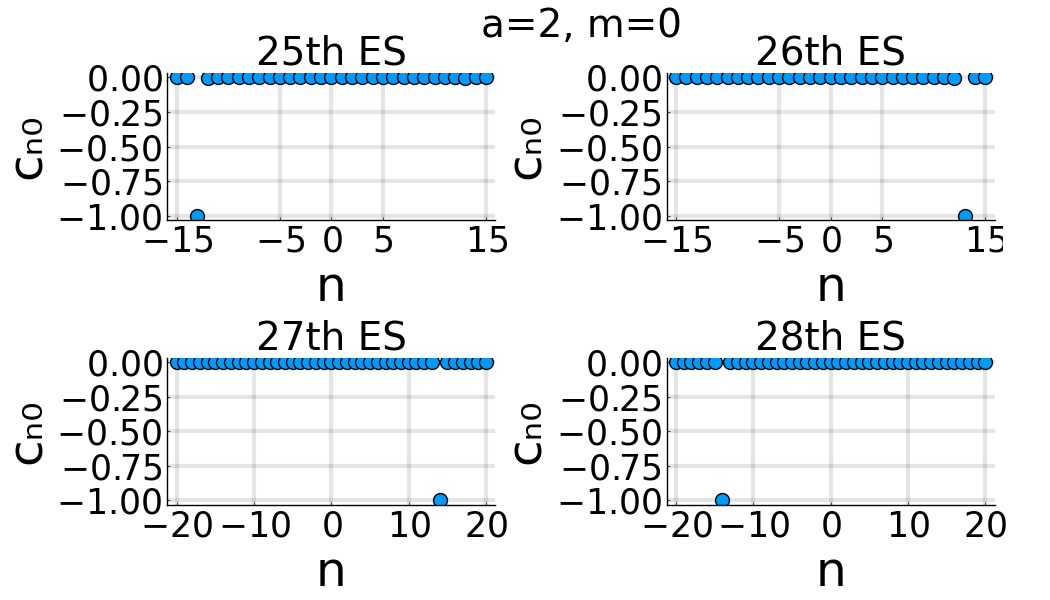}
	\hspace{1cm}
	\includegraphics[width=7cm]{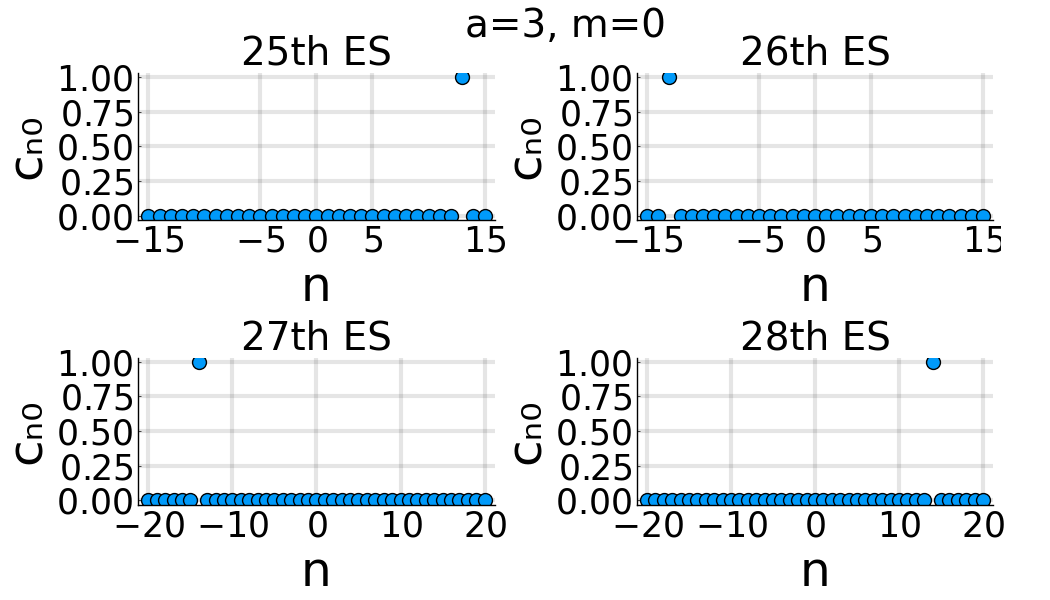}
	\caption{The coefficients of the expansion of the energy eigenvectors in the basis formed by the functions $\cF^{(\Lambda)}_{n,m} (\theta, \phi)$, for $m=0$, $a = 2$ (left) and $a = 3$ (right).}
	\label{fig_vectors25-28_m0}
\end{figure}

\begin{figure}[t]
	\centering
	\includegraphics[width=7cm]{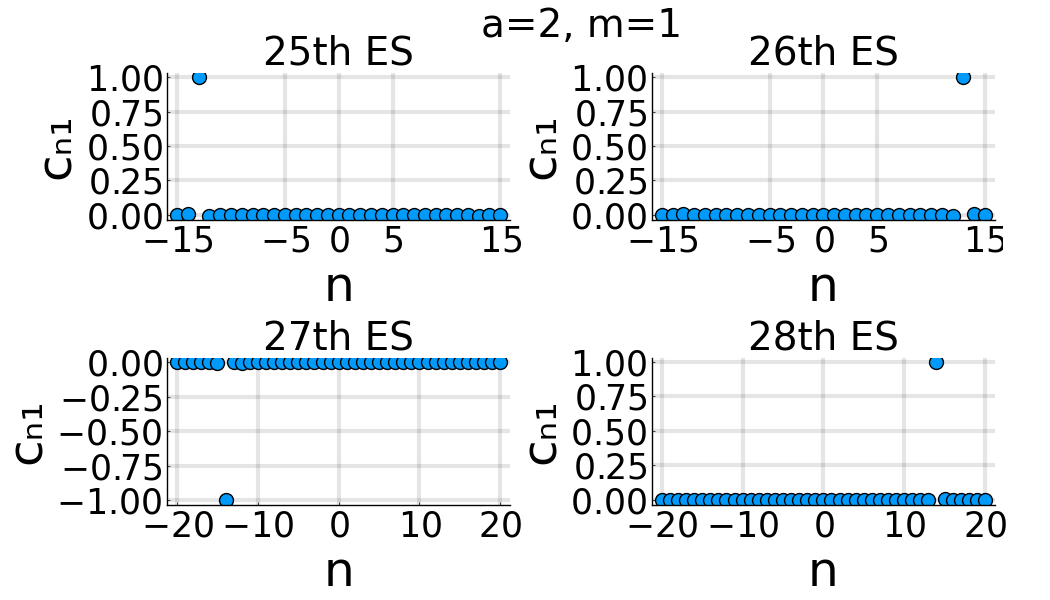}
	\hspace{1cm}
	\includegraphics[width=7cm]{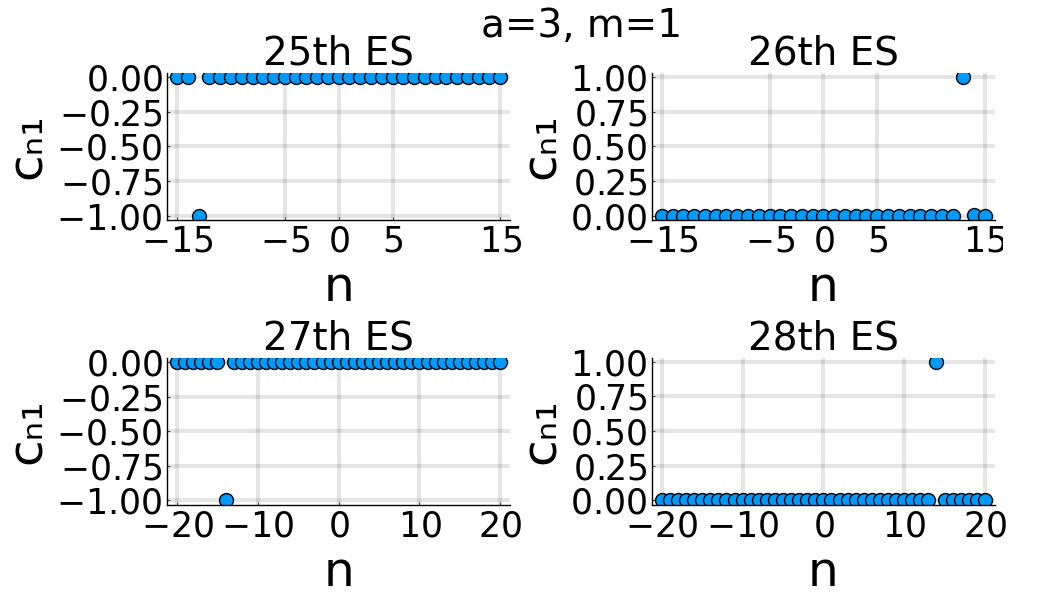}
	\caption{The coefficients of the expansion of the energy eigenvectors in the basis formed by the functions $\cF^{(\Lambda)}_{n,m} (\theta, \phi)$, for $m=1$, $a = 2$ (left) and $a = 3$ (right).}
	\label{fig_vectors25-28_m1}
\end{figure}

To understand the behavior at higher energy levels, we plot in Figs.~\ref{fig_vectors25-28_m0} and \ref{fig_vectors25-28_m1} the coefficients of the expansion of the Hamiltonian eigenstates corresponding to higher eigenvalues.
We observe that at higher energies only one of the coefficients is significantly different from zero, meaning that the energy and the momentum $\hat{p}^{(\theta)}$ have approximately the same eigenfunctions, that is, they (approximately) commute.
From Eq.~(\ref{H_elms_matr}) we understand this approximate ``commutation'' relation because while the diagonal elements increase quadratically with the excitation energy, the non-diagonal elements are independent of it and, therefore, they may be neglected at higher energies.

\section{Thermodynamics of particles on the torus} \label{sec_thermo}

Using the results of Section~\ref{sec_eign_val_tor}, we can calculate the statistics of ideal particles on the torus.
We calculate the partition function as the trace over multi-particle states of symmetrized (for bosons) or antisymmetrized (for fermions) products of eigenfunctions $\cF^{(\Lambda)}_{n_1,m_1}$.
Generally, the partition function is
\begin{eqnarray}
	\cZ &\equiv& \Tr \left\{e^{-\beta(\hat{\cH}^{(2D)} - \mu\hat{\cN})}\right\} ,
	\label{def_cZ}
\end{eqnarray}
where $\hat{\cN}$ is the particle number operator, $\mu$ is the chemical potential, $\beta \equiv 1/(k_BT)$, $T$ is the inverse temperature, and $k_B$ is Boltzmann's constant.
We ignore $\hat{\cH^{(q)}}$, since this contributes only by a constant, which may be incorporated into $\mu$.
Then, the logarithm of the partition function may be written in a compact form for both, bosons ($-$) and fermions ($+$), as
\begin{eqnarray}
	\ln \cZ^{(\pm)} &=& \pm \sum_{n,m} \ln \left[1 \pm e^{-\beta(\epsilon_{nm} - \mu)}\right]
	\label{lnZ_pm}
\end{eqnarray}
where we used~(\ref{H_diag}) to introduce the short-hand notation
\begin{equation}
	\epsilon_{nm} \equiv \frac{\hbar^2}{2m_p} \frac{a^2}{R^2}
	\left[ n^{2} + \frac{a}{\left(a^{2}-1\right)^{\frac{3}{2}}} m^{2} \right]
	\label{def_eps_nm}
\end{equation}
and the constant factors from $\langle \cF^{(\Lambda)}_{n,m} | \hat{\cH}^{(2D)} | \cF^{(\Lambda)}_{n,m} \rangle$ have also been assimilated into $\mu$.

If $\frac{\hbar^2}{2m_p} \frac{a^2}{R^2} \beta \ll 1$, then we can transform the summations in~(\ref{lnZ_pm}) into integrals and we obtain a 2D quasi-continuous gas of ideal massive particles of constant density of states.
Then, the expression~(\ref{lnZ_pm}) may be written as
\begin{eqnarray}
	\ln \cZ^{(\pm)} &=& \pm k_BT \int_{-\infty}^{\infty} \frac{dk_n}{\sqrt{\frac{\hbar^2}{2m_p} \frac{a^2}{R^2}}} \int_{-\infty}^{\infty} \frac{dk_m}{\sqrt{\frac{\hbar^2}{2m_p} \frac{a^2}{R^2} \frac{a}{\left(a^{2}-1\right)^{\frac{3}{2}}}}} \ln \left[1 \pm e^{-\beta(\kappa_n^2+\kappa_m^2 - \mu)}\right]
	\nonumber \\
%	%
%	&=& \pm 2\pi k_BT \frac{2m_p}{\hbar^2} \frac{R^2}{a^2} \sqrt{\frac{\left(a^{2}-1\right)^{\frac{3}{2}}}{a}}\int_{0}^{\infty} \kappa d\kappa \ln \left[1 \pm e^{-\beta(\kappa^2 - \mu)}\right]
%	\nonumber \\
%	%
%	&=& \pm \pi k_BT \frac{2m_p}{\hbar^2} \frac{R^2}{a^2} \sqrt{\frac{\left(a^{2}-1\right)^{\frac{3}{2}}}{a}}\int_{0}^{\infty} d\epsilon \ln \left[1 \pm e^{-\beta(\epsilon - \mu)}\right]
%	\nonumber \\
%	%
	&=& \frac{2 \pi m_p}{\hbar^2} \frac{R^2}{a^2} \sqrt{\frac{\left(a^{2}-1\right)^{\frac{3}{2}}}{a}}\int_{0}^{\infty} \frac{\epsilon d\epsilon}{e^{-\beta(\epsilon - \mu)} \pm 1}
	\label{lnZ_int}
\end{eqnarray}
and all the thermodynamics follows.

On the other hand, if we have fermions at very low temperature ($\frac{\hbar^2}{2m_p} \frac{a^2}{R^2} \beta \gg 1$), the Hamiltonian eigenstates are occupied one-by-one, in the order of increasing energy.
As explained in Section~\ref{sec_eign_val_tor}, at lower energies, the expectation value of the toroidal dipole operator are (almost) zero, but above a certain energy, the expectation value deviates from zero quite abruptly, producing an overall toroidal dipole moment of the system.

\section{Conclusions} \label{sec_conclusions}

We developed here a quantum formalism for calculating the properties of a particle on a quasi-2D domain folded into a shape with cylindrical symmetry.
We used a set of curvilinear coordinates adapted to our problem and expressed the relevant operators (momentum, Hamiltonian, and toroidal dipole) in terms of these coordinates.
We split the projection $\hat{T}_3$ of the toroidal dipole operator into two components and showed that the physically relevant one is self-adjoing and, therefore, represents an observable.
This is eventually the first time when the toroidal dipole is associated with a quantum observable.
We applied the formalism to a torus and obtained the eigenvalues and eigenfunctions for these operators (Eqs.~\ref{eigenfcts_pl_L3}, \ref{q_cond_tor}, Figs.~\ref{fig_gs_1_2_3_en_lev}, \ref{fig_states_m0_m1_m2_m3}, \ref{fig_tor10001}).
We also calculated the expectation value of the toroidal dipole operator on the Hamiltonian eigenstates and observed a crossover between states of zero toroidal dipole to states with non-zero toroidal dipole, as the energy increases (Fig.~\ref{fig_Tor_exps}) and explained this crossover by analyzing the coefficients of the expansion of the energy eigenstates in terms of momentum eigenstates (Figs.~\ref{fig_vectors1-5_m0}, \ref{fig_vectors1-5_m1}, \ref{fig_vectors25-28_m0}, \ref{fig_vectors25-28_m1}).
We also calculated the thermodynamics of a system of identical particles on a torus.
At higher temperatures, the system is equivalent to a quasi-2D system (as expected), whereas in the low temperature limit, a Fermi system becomes degenerate and, when the Fermi level exceeds the crossover energy, the system exhibits toroidal polarizability alternating with the number of particles.

In Appendix~\ref{sec_app_herm} we explicitly show the hermiticity of the toroidal dipole operator written in the curvilinear coordinates.

\section{Acknowledgments} \label{sec_ack}

%This work has been financially supported by UEFISCDI project PN-19060101/2019 and the ELI-RO contract 81-44 /2020.
This work has been financially supported by the ELI-RO contract 81-44 /2020.
Travel support from Romania-JINR collaboration projects positions 19, 22, Order 366/11.05.2021 is gratefully acknowledged.

\appendix

\section{The explicit proof of the hermiticity of the $\hat{T}_3$ operator in the $(l,q,\phi)$ coordinates} \label{sec_app_herm}

%To check the formula~(\ref{T_3tot}), we prove here the  hermiticity of the $\hat{T}_3$ in the $(l,q,\phi)$ coordinates
We prove here the  hermiticity of the $\hat{T}_3$ in the $(l,q,\phi)$ coordinates, to check the formula~(\ref{T_3tot}) and introduce some notations.
The expectation value of $\hat{T}_3$ on an arbitrary state $\Psi$ is
\begin{eqnarray}
	\langle \Psi | \hat{T}_3 | \Psi \rangle &=& \frac{-i\hbar}{10 m_p} \int_0^{2\pi}d\phi \int_0^L dl \int_0^{q_{max}} dq \rho h_l \Psi^*(l,q) \left[
	z\rho \,\hat{\brho} - \left( 2\rho^2 + z^2 \right) \,\hat{\bz} \right] \cdot \left(
	\frac{\hat{\bl}}{h_l} \frac{\partial}{\partial l}
	+ \hat{\bq}\frac{\partial}{\partial q}
	\right) \Psi(l,q) ,
	\label{T3_lq}
\end{eqnarray}
Integrating by parts Eq.~(\ref{T3_lq}), we obtain
\begin{subequations} \label{T3_lq_terms}
\begin{eqnarray}
\langle \Psi | \hat{T}_3 | \Psi \rangle &=& \frac{-i\hbar}{10 m_p} \int_0^{2\pi}d\phi \int_0^L dl \int_0^{q_{max}} dq \frac{\partial}{\partial l} \left\{ \rho \left[
z\rho \,\hat{\brho} - \left( 2\rho^2 + z^2 \right) \,\hat{\bz} \right] \cdot \hat{\bl} \, \Psi^*(l,q) \Psi(l,q) \right\} \label{T3_lq_a} \\
&& - \frac{i\hbar}{10 m_p} \int_0^{2\pi}d\phi \int_0^L dl \int_0^{q_{max}} dq \frac{\partial}{\partial q} \left\{ \rho h_l \left[
z\rho \,\hat{\brho} - \left( 2\rho^2 + z^2 \right) \,\hat{\bz} \right] \cdot \hat{\bq} \, \Psi^*(l,q) \Psi(l,q) \right\} \label{T3_lq_b} \\
&& + \frac{i\hbar}{10 m_p} \int_0^{2\pi}d\phi \int_0^L dl \int_0^{q_{max}} dq \frac{\partial}{\partial l} \left\{ \rho \left[
z\rho \,\hat{\brho} - \left( 2\rho^2 + z^2 \right) \,\hat{\bz} \right] \cdot \hat{\bl} \right\} \, \Psi^*(l,q) \Psi(l,q) \label{T3_lq_c} \\
&& + \frac{i\hbar}{10 m_p} \int_0^{2\pi}d\phi \int_0^L dl \int_0^{q_{max}} dq \frac{\partial}{\partial q} \left\{ \rho h_l \left[
z\rho \,\hat{\brho} - \left( 2\rho^2 + z^2 \right) \,\hat{\bz} \right] \cdot \hat{\bq} \right\} \, \Psi^*(l,q) \Psi(l,q) \label{T3_lq_d} \\
&& + \frac{i\hbar}{10 m_p} \int_0^{2\pi}d\phi \int_0^L dl \int_0^{q_{max}} dq \rho h_l \left[
z\rho \,\hat{\brho} - \left( 2\rho^2 + z^2 \right) \,\hat{\bz} \right] \cdot \left[
\hat{\bl} \Psi(l,q) \frac{1}{h_l} \frac{\partial}{\partial l} \Psi^*(l,q)
+ \hat{\bq} \Psi(l,q) \frac{\partial}{\partial q} \Psi^*(l,q)
\right] . \nonumber
\end{eqnarray}
\end{subequations}
The integrals in the lines (\ref{T3_lq_a}) and (\ref{T3_lq_b}) are zero by periodic (in $l$) or Dirichlet (in $q$) boundary conditions.
From the lines (\ref{T3_lq_c}) and (\ref{T3_lq_c}) we calculate
\begin{subequations} \label{calc_Tl_Tq}
\begin{eqnarray}
	T_l(l,q) &\equiv& \frac{\partial}{\partial l} \left\{ \rho \left[
	z\rho \,\hat{\brho} \cdot \hat{\bl} - \left( 2\rho^2 + z^2 \right) \,\hat{\bz} \cdot \hat{\bl} \right] \right\}
	=
	\frac{\partial}{\partial l} \left\{ z\rho^2 \,\hat{\brho} \cdot \hat{\bl} - 2\rho^3 \,\hat{\bz} \cdot \hat{\bl} - z^2 \rho \,\hat{\bz} \cdot \hat{\bl} \right\}
	\nonumber \\
	&=&
	h_l \left\{ \rho^2 \,(\hat{\brho} \cdot \hat{\bl}) \,(\hat{\bz} \cdot \hat{\bl}) + 2z\rho \,(\hat{\brho} \cdot \hat{\bl})^2 - 6\rho^2 \,(\hat{\brho} \cdot \hat{\bl}) \,(\hat{\bz} \cdot \hat{\bl}) - 2 z \rho \,(\hat{\bz} \cdot \hat{\bl})^2 - z^2 \, (\hat{\brho} \cdot \hat{\bl})\, (\hat{\bz} \cdot \hat{\bl}) \right\}
	\nonumber \\
	&& + \rho h_l \left[
	z\rho \,\hat{\brho} - \left( 2\rho^2 + z^2 \right) \,\hat{\bz} \right] \cdot \frac{\partial \hat{\bl}}{\partial l}
	=
	h_l \left\{ 2z\rho \, [ (\hat{\brho} \cdot \hat{\bl})^2 - (\hat{\bz} \cdot \hat{\bl})^2 ] - [5\rho^2 + z^2] \, (\hat{\brho} \cdot \hat{\bl})\, (\hat{\bz} \cdot \hat{\bl}) \right\}
	\nonumber \\
	&& + \rho h_l \left[
	z\rho \,\hat{\brho} - \left( 2\rho^2 + z^2 \right) \,\hat{\bz} \right] \cdot \frac{\partial \hat{\bl}}{\partial l} ,
	\label{calc_Tl} \\
	%%%%%%%%%%%%%%%%%%%%%%%
	T_q(l,q) &\equiv& \frac{\partial}{\partial q} \left\{ \rho h_l \left[
	z\rho \,\hat{\brho} \cdot \hat{\bq} - \left( 2\rho^2 + z^2 \right) \,\hat{\bz} \cdot \hat{\bq} \right] \right\}
	=
	h_l \frac{\partial}{\partial q} \left\{ z\rho^2 \,\hat{\brho} \cdot \hat{\bq} - 2\rho^3 \,\hat{\bz} \cdot \hat{\bq} - z^2 \rho \,\hat{\bz} \cdot \hat{\bq} \right\}
	\nonumber \\
	&& + \rho \left[
	z\rho \,\hat{\brho} \cdot \hat{\bq} - \left( 2\rho^2 + z^2 \right) \,\hat{\bz} \cdot \hat{\bq} \right] \frac{\partial h_l}{\partial q}
	=
	h_l \left\{ 2z\rho \, [ (\hat{\brho} \cdot \hat{\bq})^2 - (\hat{\bz} \cdot \hat{\bq})^2 ] - [5\rho^2 + z^2] \, (\hat{\brho} \cdot \hat{\bq})\, (\hat{\bz} \cdot \hat{\bq}) \right\}
	\nonumber \\
	&& + \rho \left[
	z\rho \,\hat{\brho} \cdot \hat{\bq} - \left( 2\rho^2 + z^2 \right) \,\hat{\bz} \cdot \hat{\bq} \right] \frac{\partial h_l}{\partial q} .
	\label{calc_Tq}
\end{eqnarray}
From Fig.~\ref{fig_contour} we see that $\hat{\brho} \cdot \hat{\bl} \equiv \cos\alpha = - \hat{\bz} \cdot \hat{\bq}$ and $\hat{\bz} \cdot \hat{\bl} = \sin\alpha = \hat{\brho} \cdot \hat{\bq}$ and using these relations we see that
\begin{equation}
	T_l + T_q = \rho \left\{ h_l \left[
	z\rho \,\hat{\brho} - \left( 2\rho^2 + z^2 \right) \,\hat{\bz} \right] \cdot \frac{\partial \hat{\bl}}{\partial l}
	+ \left[
	z\rho \,\hat{\brho} - \left( 2\rho^2 + z^2 \right) \,\hat{\bz} \right] \cdot \hat{\bq} \frac{\partial h_l}{\partial q} \right\} .
	\label{calc_TqpTl}
\end{equation}
\end{subequations}
From Fig.~\ref{fig_angles} we see that $|\partial \hat{\bl}/\partial l| = 1/|r(l)| = |\partial h_l / \partial q|$.
Using the signs rules explained in Fig.~\ref{fig_angles}, we observe that
\begin{equation}
	T_l + T_q = 0 , \label{Tl_Tq_0}
\end{equation}
which, together with Eqs.~(\ref{T3_lq}) and (\ref{T3_lq_terms}), implies that
\begin{eqnarray}
&& \langle \Psi | \hat{T}_3 | \Psi \rangle = \frac{-i\hbar}{10 m_p} \int_0^{2\pi}d\phi \int_0^L dl \int_0^{q_{max}} dq \rho h_l \Psi^*(l,q) \left[
z\rho \,\hat{\brho} - \left( 2\rho^2 + z^2 \right) \,\hat{\bz} \right] \cdot \left(
\frac{\hat{\bl}}{h_l} \frac{\partial}{\partial l}
+ \hat{\bq}\frac{\partial}{\partial q}
\right) \Psi(l,q) \nonumber \\
&& = \frac{-i\hbar}{10 m_p} \int_0^{2\pi}d\phi \int_0^L dl \int_0^{q_{max}} dq \rho h_l \Psi(l,q) \left[
z\rho \,\hat{\brho} - \left( 2\rho^2 + z^2 \right) \,\hat{\bz} \right] \cdot \left(
\frac{\hat{\bl}}{h_l} \frac{\partial}{\partial l}
+ \hat{\bq}\frac{\partial}{\partial q}
\right) \Psi^*(l,q)
\equiv \langle \Psi | \hat{T}_3 | \Psi \rangle^* ,
\label{T3_herm}
\end{eqnarray}
so, $\hat{T}_3$, expressed in the coordinates $(l,q,\phi)$, is hermitian, as expected.

%\bibliography{C:/Users/drago/DARKMAT-LASER Dropbox/Documentation/general}
%\bibliography{/home/dragos/general}
% \bibliography{/media/sf_Share/Dropbox/general}
%\bibliographystyle{unsrt}

\end{document}